\documentclass{emulateapj}

\usepackage{amsfonts}
\usepackage{amssymb}
\usepackage{amsmath,amsthm}
\usepackage{enumerate}
\usepackage{amsmath, amssymb, mathrsfs}
\bibpunct{(}{)}{;}{a}{}{,}

\shorttitle{High-Resolution Spectroscopy of UMa\,II and ComBer}
\shortauthors{Frebel et al.}

\begin{document}
\title{High-Resolution Spectroscopy of Extremely Metal-Poor Stars \\in the Least
  Evolved Galaxies: Ursa Major\,II and Coma Berenices\altaffilmark{1}}

\author{
Anna Frebel\altaffilmark{2,3},
Joshua D. Simon\altaffilmark{4},
Marla Geha\altaffilmark{5},
Beth Willman\altaffilmark{6}}

\altaffiltext{1}{Based on observations obtained at the W. M. Keck
Observatory, which is operated jointly by the California Institute of
Technology and the University of California, and the National
Aeronautics and Space Administration.}

\altaffiltext{2}{McDonald Observatory, The University of Texas at
Austin, Austin, TX 78712}

\altaffiltext{3}{Harvard-Smithsonian Center for
  Astrophysics, Cambridge, MA 02138; afrebel@cfa.harvard.edu}

\altaffiltext{4}{Observatories of the Carnegie Institution of
  Washington, Pasadena, CA 91101; jsimon@ociw.edu}

\altaffiltext{5}{Astronomy Department, Yale University, New Haven, CT
06520; marla.geha@yale.edu}

\altaffiltext{6}{Haverford College, Haverford, PA 19041;
  bwillman@haverford.edu}

\begin{abstract}
  We present Keck/HIRES spectra of six metal-poor stars in two of
  the ultra-faint dwarf galaxies orbiting the Milky Way, Ursa
  Major\,II and Coma Berenices. These observations include the first
  high-resolution spectroscopic observations of extremely metal-poor
  stars ($\mbox{[Fe/H]}<-3.0$) stars not belonging to the Milky Way
  (MW) halo field star population.  We obtain abundance measurements
  and upper limits for 26 elements between carbon and europium.  The
  entire sample of stars spans a range of $-3.2<\mbox{[Fe/H]}<-2.3$,
  and we confirm that each galaxy contains a large intrinsic spread of
  Fe abundances. A comparison with MW halo stars of similar
  metallicities reveals substantial agreement between the abundance
  patterns of the ultra-faint dwarf galaxies and the MW halo for the
  light, $\alpha$ and iron-peak elements (C to Zn).  This agreement
  contrasts with the results of earlier studies of more metal-rich
  stars ($-2.5\lesssim\mbox{[Fe/H]}\lesssim-1.0$) in more luminous
  dwarf spheroidal galaxies (dSphs), which found significant abundance
  discrepancies with respect to the MW halo data.  The abundances of
  neutron-capture elements (Sr to Eu) in the ultra-faint dwarf
  galaxies are extremely low, consistent with the most metal-poor halo
  stars, but not with the typical halo abundance pattern at
  $\mbox{[Fe/H]}\gtrsim-3.0$.  Our results are broadly consistent with
  a galaxy formation model that predicts that massive dwarf galaxies
  are the source of the metal-rich component ($\mbox{[Fe/H]}>-2.5$) of
  the MW halo, but we also suggest that the faintest known dwarfs may
  be the primary contributors to the metal-poor end of the MW halo
  metallicity distribution.
 \end{abstract}

\keywords{early universe --- galaxies: dwarf --- Galaxy: halo ---
Local Group --- stars: abundances --- stars: Population II }

\section{Introduction}
Dwarf spheroidal (dSph) galaxies are among the most metal-poor stellar
systems in the local universe \citep{mateo98a}, and the recently
discovered ``ultra-faint'' ($M_{V}>\sim8$) dwarf galaxies
\citep{willman05b,willman05a,zucker06a,zucker06b,
  belokurov06a,belokurov06b,sakamoto06a, irwin07a,
  walsh07a,belokurov08} are the least chemically enriched systems yet
found \citep{munoz06a,SG07,kirby08,geha08a}.  The mean metallicity
[Fe/H]\footnote{Throughout this paper, we assume that the Fe abundance
  traces the overall metallicity Z of a star. We thus use the terms
  metallicity, Fe abundance and [Fe/H] interchangeably, where
  \mbox{[A/B]}$ = \log(N_{\rm A}/N_{\rm B}) - \log(N_{\rm A}/N_{\rm
    B})_\odot$ for the number N of atoms of element A and B.} of the
twelve Milky Way ultra-faint dwarf galaxies observed so far is
$\mbox{[Fe/H]} = -2.3$ \citep{kirby08}, and the most metal-poor of
these have lower metallicities than any known globular cluster
\citep{harris97a}.  The ultra-faint dwarf galaxies are highly dark
matter-dominated \citep{martin07a,SG07,strigari08a,geha08a} and lie
on the extension of the metallicity-luminosity relationship and other
scaling relations established by brighter dSph galaxies
\citep{kirby08, penarrubia08}.  Thus, these objects appear to
represent the extreme limit of the galaxy formation process.

Detailed chemical abundance measurements of individual stars in dwarf
galaxies can provide a unique window into how star formation and
chemical enrichment proceeded in the early universe.  Such ``stellar
archaeology'' is a powerful tool for recovering the chemical
composition of the stellar birth cloud and revealing how the
star-forming gas was enriched by previous generation(s) of stars.  The
very low Fe abundances seen in the ultra-faint dwarf galaxies suggest that
perhaps only one or a few generations of star formation occurred
before the birth of the stars observed today.  These galaxies
therefore afford us an unusually clear glimpse of the nucleosynthetic
products of some of the first stars.

The past decade has provided a wealth of new information about
chemical abundances in dSphs in general (e.g., \citealt{shetrone98,
  shetrone01, shetrone03, tolstoy03, venn04, geisler05}), but the
ultra-faint dwarf galaxies were discovered too recently to have been
included in these studies.  Two of the major results from work on the
brighter dSphs were that the abundance patterns of dSph stars (most
notably the [$\alpha$/Fe] ratios) differ significantly from those seen
in the stellar halo of the Milky Way (\citealt{venn04}, and references
therein) and that the dSphs seemed to lack the extremely metal-poor
($\mbox{[Fe/H]} < -3$) stars \citep{helmi06} that are known to be
present in the Milky Way (MW) halo (e.g., \citealt{ARAA}).  Although
comparisons to the most recent unbiased determination of the halo
metallicity distribution function now indicate that the dSph
metallicity distributions may be reasonably consistent with the halo
\citep{schoerck}, it is still the case that no extremely metal-poor
stars have been identified in the brightest dSphs. 

These findings were initially interpreted as a challenge to
hierarchical formation scenarios for the Milky Way; if the stellar
halo is built up by the destruction of dwarf galaxies
\citep[e.g.,][]{sz78}, then one might naively expect stars in dwarf
galaxies to have similar properties to halo stars.  Subsequently, more
sophisticated analyses combining N-body simulations and semi-analytic
chemical evolution models demonstrated that the [$\alpha$/Fe]
discrepancy is in fact a natural byproduct of stellar halo formation
in a hierarchical universe, because the bulk of the MW halo had its
origin in satellites much more massive than the presently observed
dSphs \citep{robertson05,bullock05a,font06a,johnston08a}.  However,
this does not solve the mystery of the missing extremely metal-poor
(EMP) stars in dSphs; such stars are observed in the halo, so they
must have come from somewhere.  This remaining problem, plus the
recent discovery of EMP stars in the ultra-faint dwarf galaxies
\citep{kirby08}, strongly motivates more detailed abundance studies
of these galaxies.

Furthermore, the large majority of stars with published
high-resolution abundance measurements are relatively metal-rich ---
only 12 out of 49 have metallicities below $\mbox{[Fe/H]} = -2.0$
\citep{shetrone98, shetrone01, shetrone03, sadakane04, geisler05,
koch08a, koch_her}.\footnote{This compilation excludes 51 stars that
have been observed in Sagittarius (Sgr), all of which have
$\mbox{[Fe/H]} > -1.6$
\citep{bonifacio00,bonifacio04,mcwilliam03,monaco,chou07}, as well as
studies of stars in the Sgr stream and Sgr and Fornax globular
clusters.}  These stars are therefore unlikely to be representative of
the earliest generations of star formation in dSphs.  The most
metal-poor dSph star observed at high resolution so far, Dra~119 in
the Draco dSph \citep{shetrone98, fulbright_rich}, has a metallicity
of $\mbox{[Fe/H]}=-2.95$.  This star shows an enhancement of
$\alpha$-elements and a lack of neutron-capture elements, hinting that
perhaps the most metal-poor components of the Milky Way halo and the
dwarf galaxies are actually similar, and the disagreements only set in
at higher metallicities.

The first high resolution spectroscopy of stars in the ultra-faint
dwarf galaxies was presented by \citet{koch_her}, who observed two
stars in the Hercules (Her) dwarf.  These stars have moderately low
metallicities ($\mbox{[Fe/H]}\sim -2$), strong enhancements of the
explosive $\alpha$-elements magnesium and oxygen, and no detected
heavy elements.  With the exception of unusual [Mg/Ca] ratios,
\citeauthor{koch_her} conclude that the abundance pattern in Her is
comparable to that of extremely low-metallicity MW halo stars --- but
quite different from the typical abundances of halo stars at
$\mbox{[Fe/H]} = -2$.

Here we present the first detailed chemical abundance measurements for
two more ultra-faint dwarf galaxies, Ursa Major\,II (UMa\,II) and Coma
Berenices (ComBer). UMa\,II ($M_V = -4.2$) and ComBer ($M_V = -4.1$)
are an order of magnitude less luminous than Her, and
medium-resolution spectra indicate that their brightest stars have
significantly lower Fe abundances than the \citet{koch_her} targets in
Hercules. Using Keck/HIRES, we have obtained high-resolution spectra
for the six brightest known member stars in these two galaxies (three
stars in each galaxy). From these data we are able to measure carbon,
iron-peak and $\alpha$-element abundances, as well as neutron-capture
species such as Ba and Sr. Two of the three stars observed in UMa\,II
have $\mbox{[Fe/H]}< -3.0$, making them the most metal-poor stars
studied with high-resolution spectroscopy that do not belong to the
Milky Way.

This paper is organized as follows.  In \S\,\ref{sec:obs} we describe
the observations and our analysis techniques.  In \S\,\ref{ab_pa} the
details of the elemental abundance determinations are presented as
well as a comparison of our metal-poor stars with MW halo stars in a
similar metallicity range.  We interpret our results within the
context of previous observations of dSphs and the hierarchical
build-up of the MW halo in \S\,\ref{sec:history} and summarize our
main findings in \S\,\ref{sec:conc}.

\begin{figure}
 \begin{center}
  \includegraphics[clip=true,width=8cm, bbllx=42, bblly=131, bburx=558,
    bbury=768]{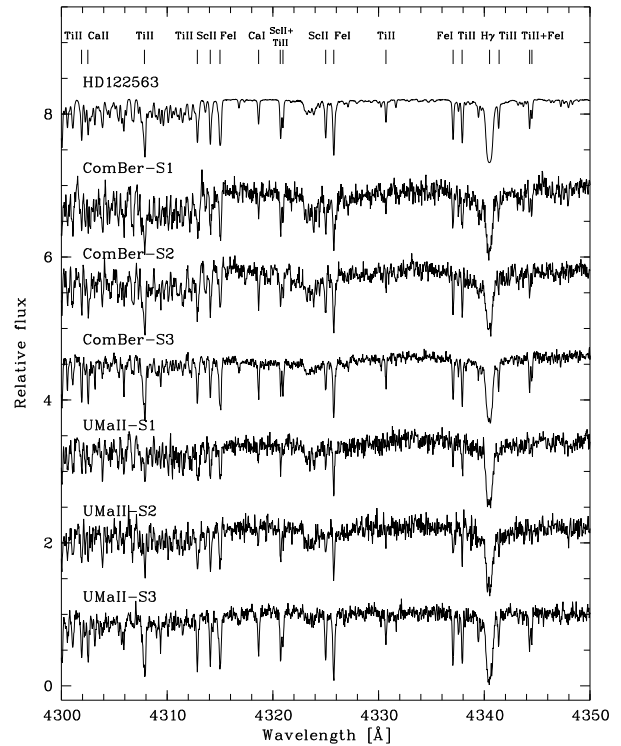}
    \figcaption{\label{hbeta_spec}Keck/HIRES spectra of our program
      stars, shown near the H\,$\gamma$ line at 4340\,{\AA}.
      Absorption lines are indicated. The CH G-band is clearly seen in
      all spectra blueward of the band head at 4313\,{\AA}.}
 \end{center}
\end{figure}

\section{Observations and Data Analysis}\label{sec:obs}
\subsection{Target Selection and Observations}
Obtaining detailed chemical abundances for stars in the ultra-faint
dwarf galaxies is challenging due to their distances (the closest is
located at 23\,kpc) and their poorly populated red giant branches.
The \citet{SG07} spectroscopic data set contains a total of nine stars
brighter than $r = 18$ that are classified as ultra-faint dwarf galaxy
members.  These stars are just bright enough to reasonably observe at
high spectral resolution with the largest available telescopes.  The
photometry of the observed stars is listed in Table~\ref{Tab:pho}.
The $r$ magnitudes and $g-r$ colors were obtained from the updated SDSS DR7
\citep{dr7}, while the $V$ magnitudes were determined from $g$ and $r$
using the conversions given by \citet{smith}.  Four of the target
stars are located in UMa\,II and three are in ComBer; our two
additional targets in Ursa~Major\,I were abandoned because of
worse than average observing conditions.  Calcium triplet metallicity
estimates for these stars indicate metallicities between
$[\mbox{Fe/H}] = -2.6$ and $-1.5$.  The \citet{kirby08} spectral
synthesis method yields even lower metallicities of $-3.0 \le
[\mbox{Fe/H}] \le -2.3$, suggesting that our targets are some of
the most metal-poor stars yet observed in dwarf galaxies.

 \begin{figure}
  \begin{center}
   \includegraphics[clip=true,width=8cm,bbllx=42, bblly=131, bburx=558,
    bbury=768]{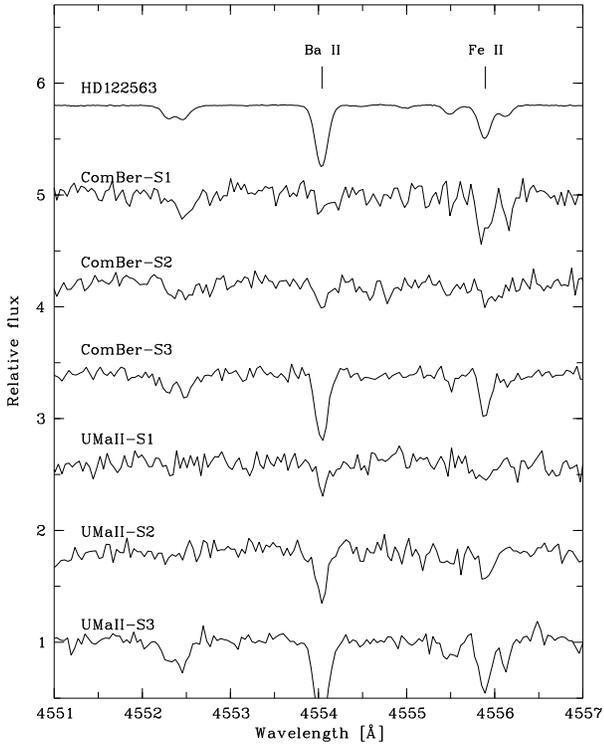}
    \figcaption{\label{ba4554_spec} Spectral region around the Ba line
      at 4554\,{\AA} for the program stars.  The
      Ba and various metallic lines are clearly detected but have
      significantly different line strengths in each spectrum,
      confirming the large spread in abundances of these elements in both UMa\,II and ComBer.}
  \end{center}
 \end{figure}

We observed the target stars with the HIRES spectrograph
\citep{vogt94} on the Keck\,I telescope on 2008 February $22-24$.
Observing conditions during the run were generally clear, with an
average seeing of 1.0\arcsec.  We used a $7.0 \times 1.15\arcsec$
slit, producing a spectral resolution of $R=37000$ over the wavelength
range from $4100-7200$\,{\AA} on the blue and green CCDs.  The red CCD
provided spectral coverage out to 8600\,{\AA}, but those data were
compromised by second-order contamination and we do not use them in
this analysis.  For one star, UMa\,II-S3, we also used a second
setting to cover bluer wavelengths from $3900-5300$\,{\AA}, 
enabling us to obtain some additional abundances for the star (see
Section~\ref{rs}). Finally, we observed the well-studied metal-poor
halo giant HD~122563 as a comparison object.  The targets, exposure
times and additional observing details are summarized in
Table~\ref{Tab:obs}.  In Figures~\ref{hbeta_spec} and
\ref{ba4554_spec} we show representative portions of the spectra of
the program stars around the H\,$\gamma$ line at 4340\,{\AA} and the
Ba line at 4554\,{\AA}.  

The echelle data were reduced with version 2.0 of the IDL software
package for HIRES developed by J.~X. Prochaska and collaborators
(Bernstein et al., in prep.).\footnote{Documentation and code for this
package can be found at http://www.ucolick.org/$\sim$xavier/HIRedux/}
The data reduction followed standard procedures, including bias
subtraction, flat-fielding, and cosmic ray rejection.  Wavelength
calibration was accomplished with ThAr comparison lamp frames taken at
the beginning and the end of each night.  After summing the available
frames for each object, the reduced and extracted spectra were
normalized using a polynomial fit to the shape of each echelle order,
excluding regions affected by absorption features.  Finally, the
overlapping echelle orders were merged together to produce the final
spectrum.

\begin{figure}
 \begin{center}
   \includegraphics[clip=true,width=8cm,bbllx=40, bblly=369,
   bburx=440, bbury=676]{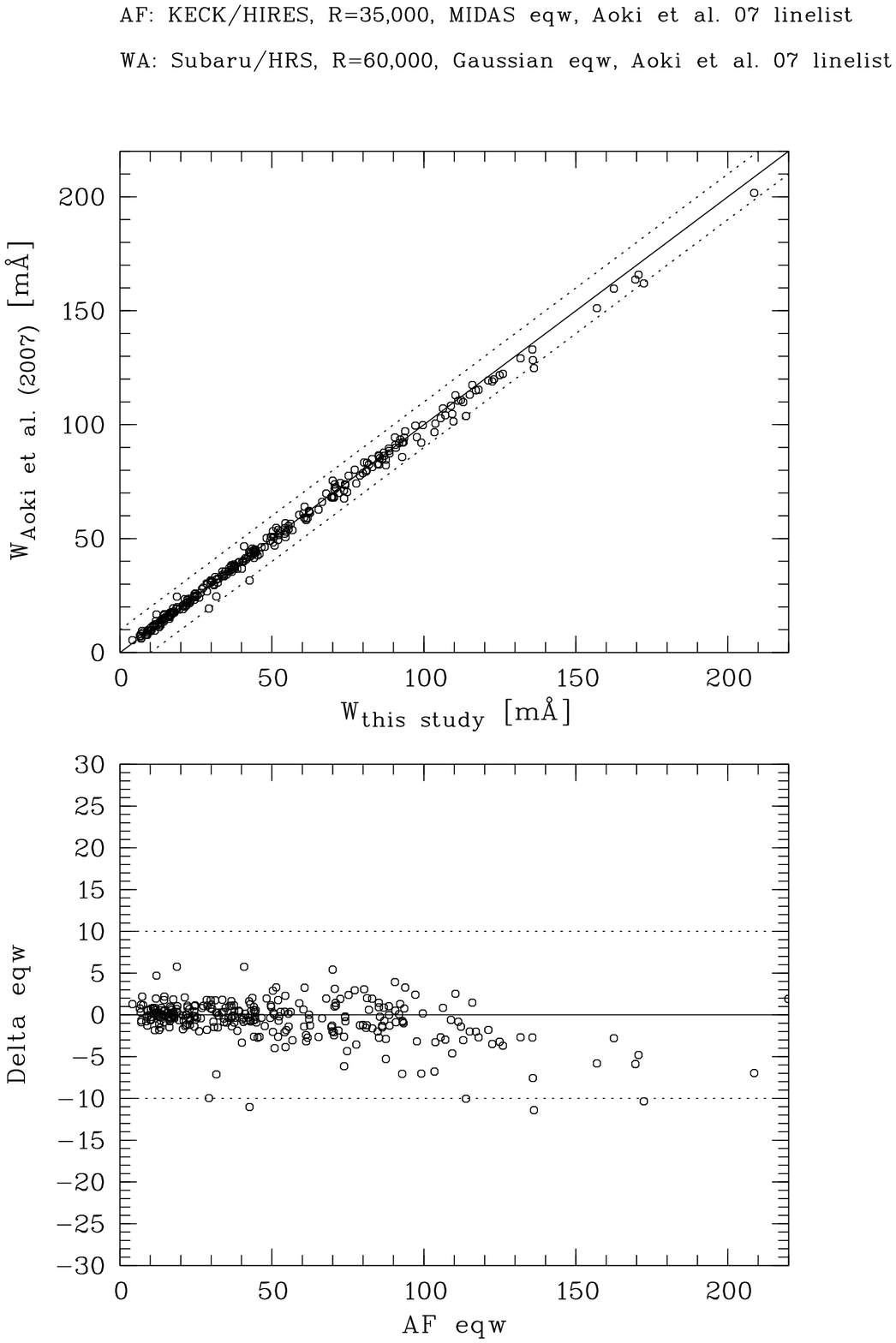} \figcaption{\label{eqw_hd122563}
     Comparison of the equivalent width measurements in our HIRES
     spectrum of HD122563 with those listed in Aoki et al.~(2007).
     The solid line indicates one-to-one correspondence, the dotted
     lines represent deviations of $\pm10$\,m{\AA} from equality
     to guide the eye.}
 \end{center}
\end{figure}

Our typical integration times of several hours per star yielded $S/N$
ratios of 25 to 30 per pixel at 5000\,{\AA}, which is sufficient for
the detection of weak spectral features.  To characterize the quality
of the data, we use the "figure of merit" introduced by
\citet{Norrisetal:2001} to compare these observations to others that
have been obtained for dwarf galaxies. In Table~\ref{Tab:merit}, we
list the figures of merit for all previous dSph observations at high
resolution.  These illustrate that even for stars fainter than 17th
magnitude it is possible to obtain high-resolution spectra of adequate
$S/N$ for detailed abundance studies in a few hours with the largest
current telescopes.

\begin{figure}
 \begin{center}
  \includegraphics[clip=true,width=8cm, bbllx=72, bblly=195,
    bburx=400, bbury=573]{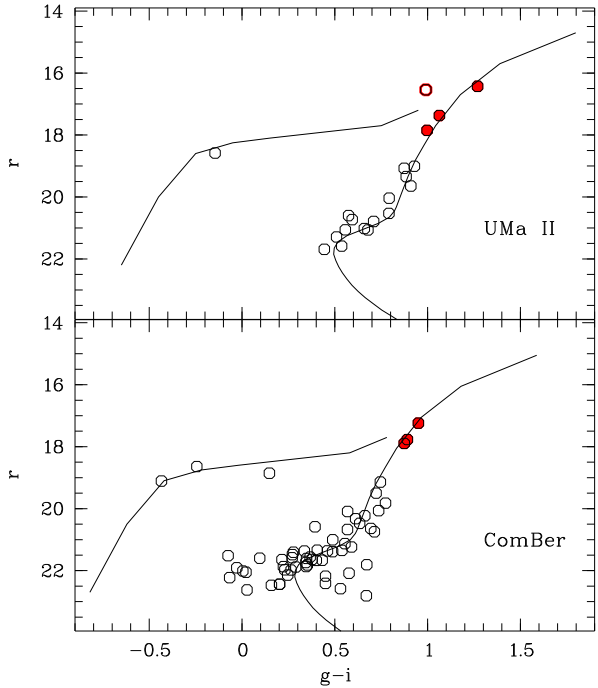} \figcaption{\label{cmd}Updated SDSS DR7
    photometry for radial velocity members of UMa\,II (top panel) and
    ComBer (bottom panel) from \citet{SG07}. Also shown are the
    isochrone of M92 and the horizontal branch of M13 (solid lines),
    both corrected for Galactic extinction and shifted to  distances
    of 32\,kpc and 44\,kpc for the two dwarf galaxies (data from
    \citealt{clem2005}).  Our high-resolution targets are shown with
    red solid circles. The red open circle refers to UMa\,II-NM, As can
    be seen, this star sits slightly off the giant branch track.}
 \end{center}
\end{figure}

During the course of our analysis, it became clear that one of the
seven observed stars, UMa\,II-NM, is a foreground MW dwarf star with a
similar velocity to UMa\,II rather than a genuine member.  The
spectroscopic surface gravity of $\log{g} = 3.5$, derived from the
usual Fe\,I-Fe\,II ionization balance argument, places the star at a
distance of only a few kpc (the distance of UMa\,II is 32\,kpc).  The
star sits slightly blueward of the UMa\,II red giant branch, where a
star descending the asymptotic giant branch (AGB) to the horizontal
branch might be located (see Figure~\ref{cmd}), but if UMa\,II-NM were a
member its the surface gravity would be $\log{g} \lesssim 2$.  Finally,
the metallicity of UMa\,II-NM of $\mbox{[Fe/H]} = -1.02$ would be
unusually high for UMa\,II, which has a mean Fe abundance more than
1\,dex lower.  We therefore discarded this star from the sample,
although we do report observational details and stellar parameters for
it in Tables~\ref{Tab:pho} and \ref{Tab:obs}.  Because this star has a
velocity relatively close to the mean velocity of UMa\,II, removing it
from the member sample does not significantly revise the systemic
velocity or velocity dispersion derived by \citet{SG07}.

\subsection{Line Measurements}\label{lines}
We use two of the strong \ion{Mg}{1}\,b lines in the green part of the
HIRES spectra as well as three other Mg lines for our radial velocity
measurements. The standard deviation of the individual line
measurements is typically $<0.5$\,km\,s$^{-1}$.  This value increases
to 0.7-$1.0$\,km\,s$^{-1}$ (usually depending on the $S/N$ level and
strength of the absorption line) after all equivalent width
measurements have been carried out based on the Mg line-derived radial
velocity correction. Consequently, the standard error is no more than
0.1\,km\,s$^{-1}$, indicating that the statistical uncertainty in our
radial velocities are very small. We measure the velocity of HD~122563
to verify that our measurements are on the correct velocity scale.  We
find a velocity of $-25.1 \pm 0.2$\,km\,s$^{-1}$ for the star, in
reasonable agreement with the velocity determined by
\citet{aoki_studiesIV} of $-26.0\pm0.2$\,km\,s$^{-1}$, as well as
other literature values. The HIRES radial velocities are slightly
offset (1 to 3\,km\,s$^{-1}$) from those obtained by \citet{SG07} from
DEIMOS $R\sim6,000$ spectra. We speculate that this offset results
from the DEIMOS velocity zeropoint determined by \citet{SG07}, which
was tied to a different set of stars, and hence should not indicate
any problems with the HIRES measurements.

For the measurements of atomic absorption lines we employ a line list
based on the compilations of \citet{aoki_studiesIV} and
\citet{ivans07}. We added the newly determined Fe\,II $gf$ values of
\citet{fe_loggf}. The molecular line data employed for CH were provided
by B. Plez (Plez, B. et al. 2009 in preparation; the latest version of the
list is described in \citealt{plez_CH} and some basic details are
given in \citealt{Hilletal:2002}). Hyperfine-structure (HFS) data for
Sc and Mn were taken from the Kurucz compilation
\citep{kurucz_lines}\footnote{http://kurucz.harvard.edu/}.

In Table~\ref{Tab:Eqw} we list the lines used and their measured
equivalent widths. Only line measurements with reduced equivalent
widths $\log (EW/\lambda)<-4.5$ were employed in the abundance
analysis of each star. The Mg\,b lines were thus excluded in several
cases since these lines are too strong and fall in the flat part of
the curve-of-growth. While we did measure the Mg\,b triplet and other lines, if
they were not used in the analysis they are not listed in
Table~\ref{Tab:Eqw}. We verify our equivalent width measurement
techniques by comparing the results for HD~122563 with the study of
\cite{aoki_studiesIV}, which covered a similar wavelength
range. Figure~\ref{eqw_hd122563} illustrates the excellent agreement
between the two data sets. For blended lines, lines with HFS, and
molecular features, we use a spectral synthesis approach. The
abundance of a given species is obtained by matching the observed
spectrum to a synthetic spectrum of known abundance.

Since the average $S/N$ ratio of the data is modest, we also
calculate what the minimum  detectable equivalent width is
for our data. We chose one example star, UMa\,II-S1, which has among
the lowest $S/N$ in the sample, has the highest temperature and the
second lowest metallicity which means that the lines are
very weak. Quantifying the minimum level in this star thus serves as
rather conservative estimate for the entire sample. Using the formula
given in \citet{Norrisetal:2001}, we estimate the approximate 3$\sigma$
detectable equivalent width to be $\sim22$\,m{\AA} for the blue and
$15$\,m{\AA} for the red part of the data (2$\sigma$ values are
$\sim17$\,m{\AA} and $10$\,m{\AA}, respectively).

A number of lines fall in the range between the 2$\sigma$ to 3$\sigma$
for this star. However, since our equivalent width measurements have
significant uncertainties from the continuum placement, we do not
discard these lines. Another reason for keeping these (mostly Fe)
lines is that weak lines near the detection limit are needed for the
determination of the microturbulent velocity.  Only five measured
lines in UMa\,II-S1 (of Ti\,I, Ti\,II, Fe\,I and two Ni\,I) have
equivalent widths less than 2$\sigma$ according to the
\citet{Norrisetal:2001} criterion (none below 1$\sigma$), but we
emphasize that ina strict statistical sense, all of thse features are
detected at the 3\,$\sigma$ level or higher. Those lines are however
found to generally yield abundances in good agreement with the
stronger lines.  In the case of Ni, only three lines could be measured
in the star and they are all very weak. Two of them have a 1.5$\sigma$
detection while third one is only at the 2.5$\sigma$ level. In the
absence of stronger Ni lines we keep the present measurements but
assign a nominal uncertainty of 0.40\,dex (based on consideration of
the measurement uncertainties). Regarding the other stars, the
situation is less severe since the $S/N$ ratio of the data is
generally better, the lines are stronger and more lines are
available. The exception is UMa\,II-S2 for which we have no detected
Ni lines, and therefore only an upper limit for the Ni abundance.

Upper limits on abundances of elements for which no lines were
detected can provide useful additional information for the
interpretation of the overall abundance patterns, and the possible
origins of the stars of interest. Based on the $S/N$ ratio in the
spectral region of the line, and employing the formula given in
\citet{o_he1327}, we derive 3\,$\sigma$ upper limits for several
elements. In Table~\ref{Tab:Eqw}, we list upper limits of a given
element for whichever line produced the tightest upper limit.

\subsection{Stellar Parameters}\label{stellpar}

\begin{figure*}
 \begin{center}
  \includegraphics[clip=true,width=16.5cm, bbllx=33, bblly=460, bburx=540,
   bbury=690]{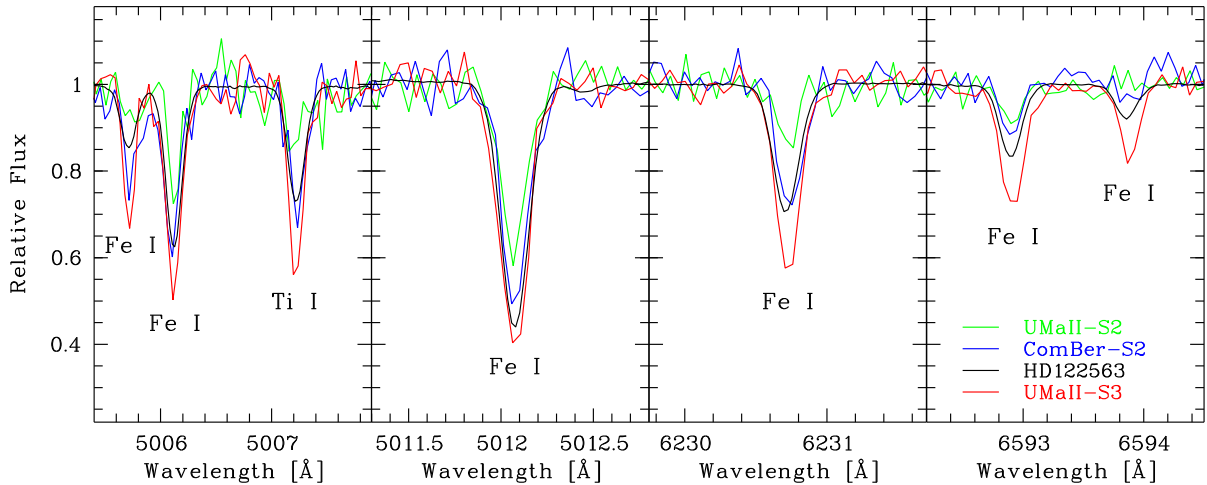} \figcaption{
     \label{felines} Several Fe\,I lines of three stars with similar
     temperatures to illustrate Fe abundance differences.  UMa\,II-S2
     (green) with $\mbox{[Fe/H]}=-3.2$, ComBer-S2 (blue) with
     $\mbox{[Fe/H]}=-2.9$, and UMa\,II-S3 with $\mbox{[Fe/H]}=-2.3$
     (red). The black line refers to HD122563, the MW halo star with
     $\mbox{[Fe/H]}=-2.8$.}
 \end{center}
\end{figure*}

\begin{figure}
 \begin{center}
   \includegraphics[clip=true,width=8cm,bbllx=40, bblly=421,
   bburx=426, bbury=628]{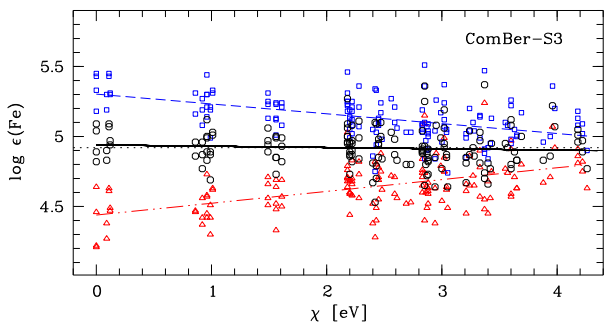} \figcaption{ \label{t_excit} Fe\,I
   abundances as a function of excitation potential, $\chi$, in
   ComBer-S3 as an example. Three different temperatures are shown:
   4800\,K (blue squares), 4600\,K (black open circles), 4400\,K (red
   triangles). The dotted line indicates the mean abundance of all Fe
   lines for the adopted temperature of 4600\,K. The dashed/solid/dot-dashed lines shows
   the corresponding fits to the data sets.}
 \end{center}
\end{figure}

\begin{figure}
 \begin{center}
   \includegraphics[clip=true,width=8cm,bbllx=67, bblly=270,
   bburx=500, bbury=576]{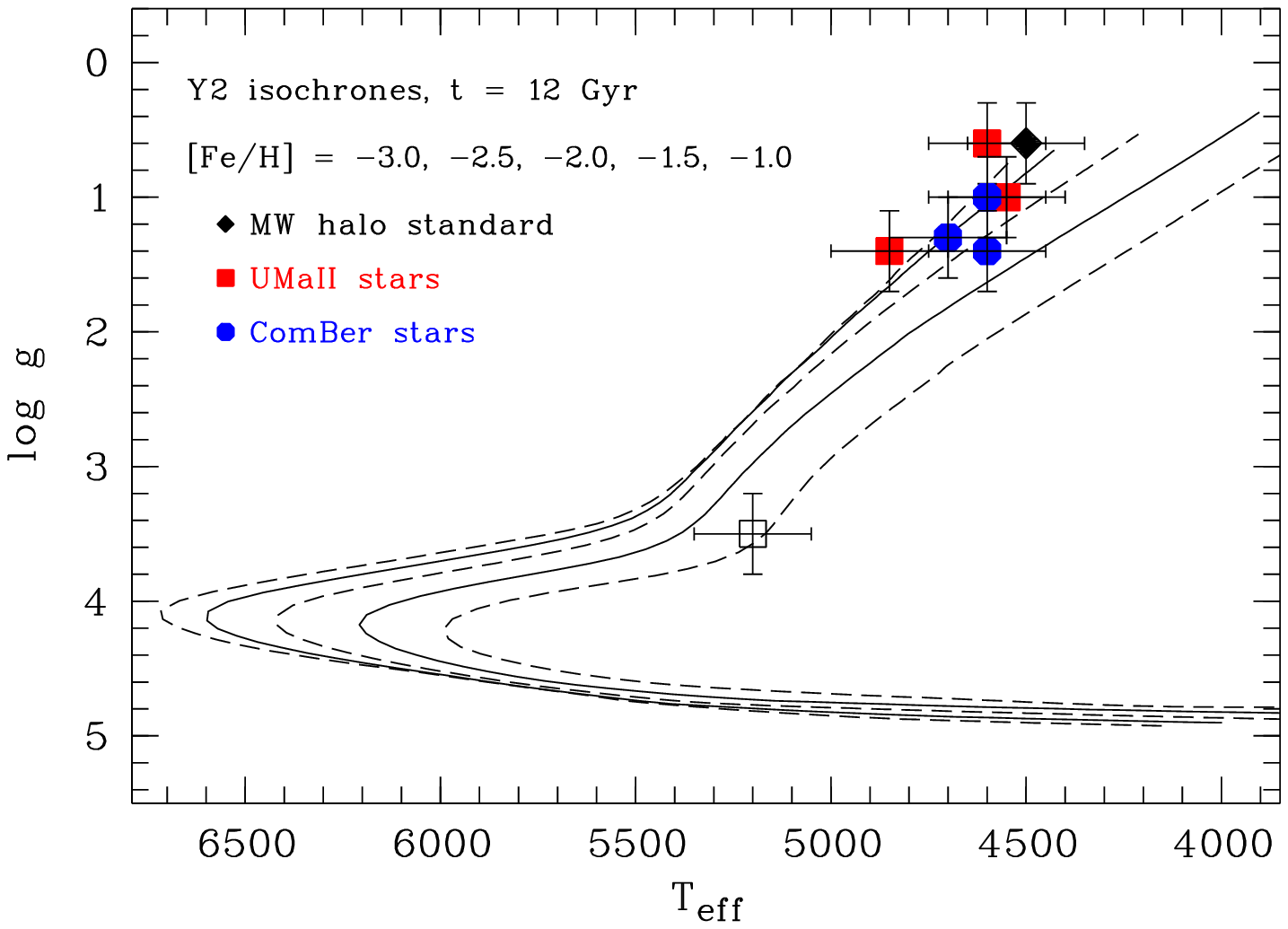} \figcaption{ \label{iso} Adopted stellar
     parameters in comparison with 10\,Gyr isochrones with
     $\mbox{[$\alpha$/Fe]}=0.4$ and metallicity ranging from
     $\mbox{[Fe/H]} = -1$ to $-3$ \citep{green, Y2_iso}.  The more
     metal-rich isochrones are shifted toward lower temperatures at a
     given surface gravity.  Red squares indicate UMa\,II stars, blue
     circles are ComBer stars. The open square shows the observed
     star that turned out to be a non-member of UMa\,II. The MW
     standard star HD122563 is marked with a black diamond.}
 \end{center}
\end{figure}

In Figure~\ref{felines} we show pieces of spectra containing a number
of \ion{Fe}{1} lines for three dwarf galaxy stars and HD~122563. All
of the stars have similar temperatures and thus allow for a simple,
visual comparison of the \ion{Fe}{1} line strengths. This quickly
illustrates the different metallicities sampled by our program
stars. It is clear that UMa\,II-S2 has much weaker Fe lines than the
other stars, demonstrating even without any analysis that this star
must be more metal-poor than HD~122563 (at $\mbox{[Fe/H]}=-2.8$).

\subsubsection{Effective temperature}
We then derive spectroscopic effective temperatures by demanding that
there be no trend of abundances with excitation potential for the
\ion{Fe}{1} lines.  As an example, Figure~\ref{t_excit} shows
\ion{Fe}{1} abundances as a function of excitation potential based on
our spectroscopically derived value for ComBer-S3. We also show Fe
abundances for temperatures of $\pm200$\,K to illustrate the
sensitivity of the method to the assumed temperature.  By varying the
temperature and comparing the derived trends to zero given the
statistical uncertainty on the slope, we determine the effective
temperature and its uncertainty. As illustrated in
Figure~\ref{t_excit}, a 200\,K change in temperature causes a strong
trend in the abundances as a function of excitation
potential. Generally, at 3\,$\sigma$ confidence we are able to determine
the temperature to within $\sim150$\,K using this technique.

The advantage of this approach over photometric temperatures is that
it is reddening-free and independent of the empirical calibrations
that are needed to convert stellar colors into effective
temperatures. Nevertheless, for completeness, we calculated
photometric temperatures from various $ugriz$ colors by using the
Yonsei-Yale isochrones \citep{Y2_iso} and the color tables of Castelli
(http://wwwuser.oat.ts.astro.it/castelli/). The differences between
the temperatures obtained from the different colors vary between
$\sim150$ to more than 300\,K for a given metallicity. While the
average temperatures agree well with our spectroscopic values (within
100\,K) for some stars, most of them agree to within 250\,K. The
spectroscopically derived temperatures are lower than photometrically
derived ones. \citet{kirby08} also calculated photometric temperatures
(also using the Yonsei-Yale isochrone, but with different color
tables; E. Kirby 2008, priv.~comm.) and in most cases, our
spectroscopic values agree with their photometric values within
200\,K. Systematic uncertainties regarding the determination method of
temperature can be estimated to be $\sim200$\,K.

\subsubsection{Surface gravity}

Using the ionization balance, i.e., demanding that \ion{Fe}{1} lines
yield the same abundance as \ion{Fe}{2} lines, we derive the surface
gravity, $\log g$, for all of the stars. Based on the standard
deviations of the averaged Fe\,I and Fe\,II abundances ($\sim0.15$ to
0.25\,dex), we estimate an uncertainty of 0.3\,dex in $\log g$. The
micro-turbulence, \mbox{v$_{\rm micr}$}, is obtained iteratively in
this process by demanding no trend of abundances with equivalent
widths. Uncertainties in this parameter are estimated to be
0.3\,km\,s$^{-1}$. Table~\ref{Tab:stellpar} lists the individual
stellar parameters. Figure~\ref{iso} shows the adopted stellar
parameters of our program stars in comparison with $\alpha$-enhanced
($\mbox{[$\alpha$/Fe]}=0.4$) 12\,Gyr isochrones \citep{green, Y2_iso}
covering a range of metallicities. Our values generally agree very
well with those of the isochrone. This also shows that our Fe line
abundances are probably not significantly affected by non-LTE effects.

\subsection{Model Atmospheres}\label{model}
Our abundance analysis utilizes 1D plane-parallel Kurucz model
atmospheres with no overshooting \citep{kurucz}. They are computed
under the assumption of local thermodynamic equilibrium (LTE). We use
the 2002 version of the MOOG synthesis code \citep{moog} for this
analysis.  Scattering in MOOG is currently treated as true
absorption. The missing implementation of a source function that sums
both absorption and scattering components (rather than treating
continuous scattering as true absorption) will be incorporated in
future MOOG versions (J.~Sobeck et al.~2009, in prep.).

In order to arrive at our final abundance ratios [X/Fe], which are
given with respect to the solar values, we employ the \citet{solar_abund}
solar abundances.  The elemental abundances for all of the target
stars are given in Tables~\ref{abundances_uma} and
\ref{abundances_comber}.

\subsection{Uncertainties}\label{unc}

\subsubsection{Measurement and Stellar Parameter Uncertainties}

To assess the level of measurement uncertainties we made use of the
fact that a number of absorption lines fall at the end of a given
order and are thus measurable twice in the two consecutive orders.  By
comparing both sets of line measurements we robustly estimate that the
average total uncertainty on our equivalent width measurements is
14\,m{\AA} for the blue lines and 9\,m{\AA} for red lines (where the
division between blue and red for this purpose is $\sim$5700\,{\AA}).
These uncertainties translate into abundance uncertainties of
$\sim0.20$\,dex. This is in good agreement with the standard
deviations of the abundances derived from individual lines of a given
element (as long as the number of lines is more than just a few), and
can thus be regarded as a general, robust estimate of the continuum
placement uncertainty (``random uncertainty'') in our derived
abundances.

\begin{figure}
 \begin{center}
   \includegraphics[clip=true,width=8cm,bbllx=67, bblly=111,
   bburx=384, bbury=683]{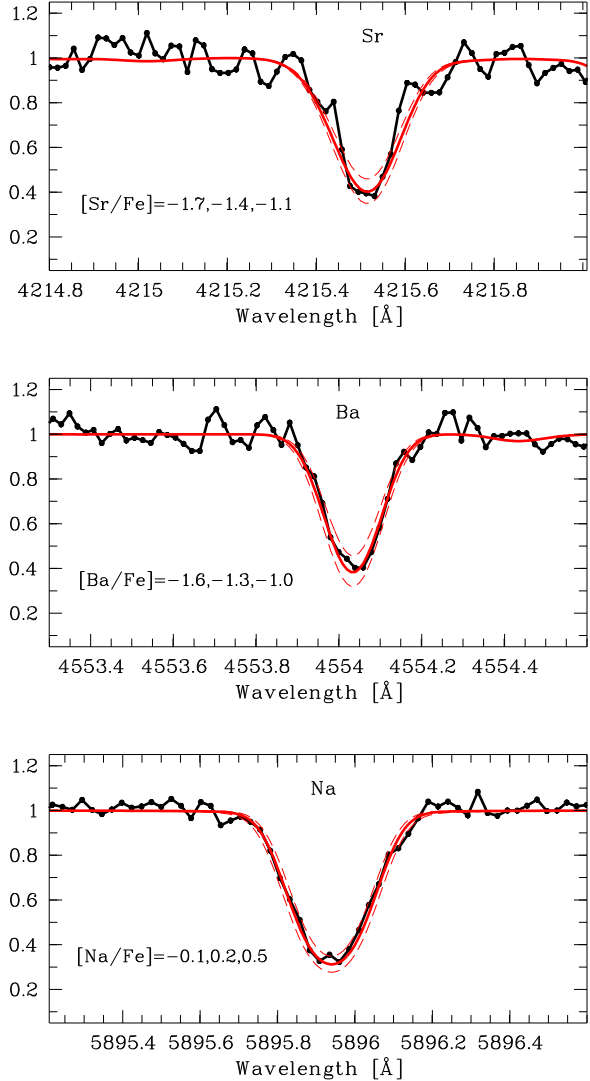} \figcaption{ \label{specs} Examples of
     strong lines (Sr, Ba, Na) used in the analysis of ComBer-S3, the
     star with the strongest Ba line (see
     Figure~\ref{ba4554_spec}). Black lines are the data, red lines
     refer to synthetic spectra of given abundances. The synthetic
     spectra demonstrate that even for lines beyond the linear part
     of the curve of growth, abundances are measurable to within
     the $\sim0.3$\,dex or better. }
 \end{center}
\end{figure}

In Figure~\ref{specs}, we show the sensitivity of the abundances
derived from strong lines near the flat part of the
curve-of-growth. From this we estimate the measurement uncertainties
of Sr, Ba and Na abundances to be 0.3\,dex. This is in agreement with
what is found from the continuum placement uncertainties and should
account for the line strengths.

For abundances of elements represented by only one line (other than
Sr, Ba, and Na) we adopt a formal uncertainty of 0.20\,dex, based on
the measurement uncertainties investigated above (these values do not
apply to HD122563 because of the higher $S/N$ level).  Despite having
two measurements for C, we also adopt 0.20\,dex for this element since
the continuum placement can be difficult for molecular bands. For many
elements where only few lines are measured the standard deviations are
unrealistically small ($<0.10$\,dex) compared with our finding
regarding the general measurement uncertainties. We thus adopt a
minimum uncertainty of 0.10\,dex in such cases.

We tested the robustness of our derived abundances by changing
one stellar parameter at a time by an amount  approximately equal
to its random uncertainty.  In Table~\ref{err} we give a summary of these
individual sources of error as well as a total uncertainty for each
element of an example star, ComBer-S3.  Taking all the sources of
errors into account, the abundances derived from atomic lines have an
average uncertainty of $\sim0.25$\,dex.

\subsubsection{Uncertainties from different log $gf$ values} 

In Section~\ref{ab_pa}, our primary comparison sample will be that
of Cayrel et al. (2004) and Francois et al. (2007), which uses
slightly different atomic data than we do.  To determine whether our
particular choice of log\,$gf$ values (as adopted from
\citealt{aoki_studiesIV}; Section~\ref{lines}) results in systematic
abundance offsets, we compare our values to those of Cayrel et al.  We
find small log\,$gf$ differences for most elements.  When more than
just a few lines are measured for each element, we estimate a constant
offset for all stars based on the number of lines in common between
our list and that of Cayrel et al.  For Mg\,I, we estimate the
required offset to be $-$0.11\,dex, for Ca\,I +0.05\,dex, for Ti\,I
+0.05\,dex, for Ti\,II $-0.09$\,dex, for Sc\,II +0.09\,dex, and for
Zn\,I $-0.04$\,dex. For Mg\,I we found larger offsets for a small
number of lines.

We find no significant systematic offset for Fe\,I (very few
individual lines have larger $gf$ differences but those are averaged
out in the final abundances), Na, Cr and Mn. No lines in common
are found for Al, Si, and Ni so we cannot derive an offset. We then
compare the abundances corrected for the log $gf$ differences as
listed above. We caution, however, that the offsets applied are based
only on the subsets of lines in common and do not reflect a more
detailed comparison of the employed $gf$ values themselves. We also
note for completeness that we recalculated the Cayrel et al. relative
abundances with the same solar abundances employed here
\citep{solar_abund}.

\subsubsection{Model Atmosphere Uncertainties}\label{model_atm_err}
Systematic uncertainties arising from the choice of model atmospheres
may add to the error budget. To test the effect of our choice of model
atmospheres, we ran a differential abundance analysis for the star
ComBer-S3 by employing a Kurucz model and the MOOG model atmosphere
code as well as a MARCS model \citep{marcs} and a corresponding code
(Uppsala LTE spectrum synthesis code ``BSYN'', version 7.05).  

 \begin{figure} 
  \begin{center}
    \includegraphics[clip=true,width=9cm,bbllx=56, bblly=105,
    bburx=396, bbury=644]{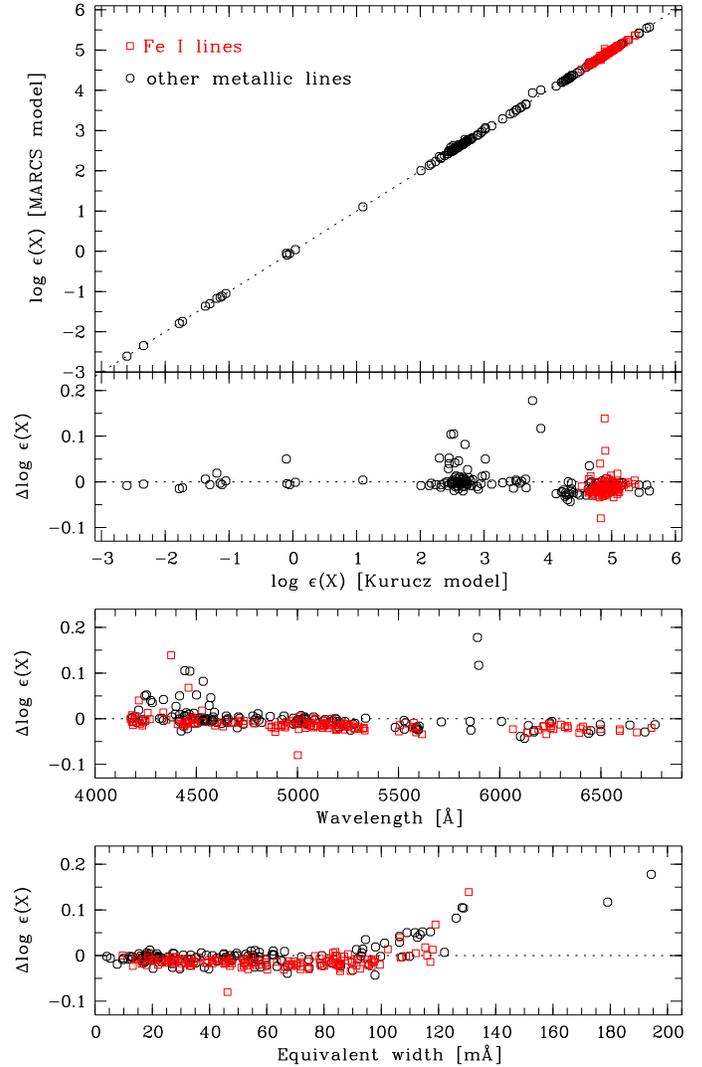} \figcaption{\label{marcs} Comparison of
    abundances of star ComBer-S3 (with stellar parameters
    \mbox{T$_{\rm eff}=4600$\,K}, $\log g = 1.0$,
    $\mbox{[Fe/H]}=-2.5$) obtained with Kurucz and MARCS model
    atmospheres. The residuals are also shown as a function of
    wavelength and equivalent width (bottom two panels).  Red squares
    indicate Fe\,I lines. Except for the strongest lines there are no
    significant differences between the two model atmospheres. }
  \end{center}
 \end{figure}

Using the BSYN code we are able to choose whether a more proper
scattering treatment (than what is used in MOOG) is ``switched
on''. This helps in quantifying various effects associated with the
different model atmospheres. We first compare the codes with no
scatter treatment in place. The difference in abundances ($\log
\epsilon (\mbox{X})$\footnote{$\epsilon ({\rm X}) = \log(N_{\rm
X}/N_{\rm H}) + 12.0$}) for individual lines are less than 0.03\,dex
for the vast majority of lines. We checked that the relative
abundances [X/Fe] are not significantly affected.  In
Figure~\ref{marcs}, we show the abundance differences as a function of
wavelength and equivalent width. As can be seen, significant deviations
($\sim0.1$ to 0.2\,dex) are only present for the strongest lines
($>120$\,m{\AA}). Since Na has only two strong lines available, this
element would be the only one affected systematically. However, since
ComBer-S3 has the strongest Na lines in the sample the effect would
likely be less pronounced in the more metal-poor stars. For the
comparison with the \citet{cayrel2004} sample in
Figure~\ref{cayrel_abundances_light} we apply a constant offset of
+0.15\,dex to all our Na abundances since the Cayrel et al. study
employed model atmospheres more closely related to the MARCS model we
used for this test.

Since \citet{cayrel2004} use a model atmosphere code that accounts for
the scattering, we also investigate what the gross effect of the
simplified treatment in MOOG would be. Using BSYN with the proper
scatter treatment, find the known dependency of abundance with
wavelength (bluer lines are more affected than red lines). However,
since we have relatively few lines with wavelengths bluer than
$\sim4500$\,{\AA} and most elements also (or exclusively) have lines
at redder wavelengths, the bias is small. Since we are using relative
abundances [X/Fe], these abundance ratios are even less affected. We
quantify the differences as follows: Mg, Ca, Ti\,II, and Cr have an
average offset over all lines of $\sim0.00$\,dex (i.e. their [X/Fe]
ratios are not affected), Ti\,I, Sc, Mn, and Zn have offsets of
$\sim+0.03$\,dex, Co and Ni have offsets of $\sim+0.05$\,dex, and Na
has $+0.20$\,dex.  Since these offsets are small we do not apply them
before plotting our abundances in
Figure~\ref{cayrel_abundances_light}. The only exception is Na for
which we apply both this and the offset determined above (for a total
of $+0.35$\,dex).

Finally, we note that while the relative abundances are only minimally
affected (except for Na), this analysis suggests that the lack of
scattering treatment in MOOG leads to an \emph{overestimate} of the Fe
abundances (both [Fe I/H] and [Fe II/H]) of our sample by 0.1\,dex.
This fact supports our finding of two stars with
$\mbox{[Fe/H]}\lesssim-3.0$ metallicities since we are adopting
slightly more conservative values.

Informed by all these tests, we conclude that the choice of model
atmosphere is a negligible source of error, but that the different
treatment of scattering can lead to small systematic offsets. For very
strong lines, and when the majority of lines for an element are
located below $\sim4300$\,{\AA} these differences are more pronounced.
Because few of our abundances rely on such blue lines (and those
measurements already have appropriately large uncertainties because of
the $S/N$), this effect is not a concern for the present study.

\subsection{Comparison with HD~122563}

One of the main objectives of this study is to compare stellar
abundances from dwarf galaxies with those of MW halo stars.  To verify
that our measurements are on the standard abundance scale, we observed
the archetypal metal-poor halo giant HD~122563 and analyzed its
spectrum in the same way as for the other stars.  The metallicity of
this star ($\mbox{[Fe/H]}\sim-2.8$) is roughly in the middle of the
range covered by our program stars, and its temperature is also very
similar.  Its stellar parameters are listed in
Table~\ref{Tab:stellpar}.  In Table~\ref{Tab:Abundances}, we list both
our [X/Fe] abundances and those of the recent \citet{aoki_studiesIV}
study (supplemented with neutron-capture abundances from
\citealt{honda06}). Considering that we derived a temperature that
differs from that of Aoki et al. (2007b) by 100\,K (Aoki et al. find
$T_{\rm{eff}}=4600$, $\log g=1.1$, $\mbox{[Fe/H]}=-2.6$ and
$v_{\rm{micr}}=2.2$; Honda et al. find $T_{\rm{eff}}=4570$, $\log
g=1.1$, $\mbox{[Fe/H]}=-2.77$ and $v_{\rm{micr}}=2.2$), the abundance
ratios we measure generally agree well.  The good agreement for most
elements lends confidence that our stellar abundance measurements for
the ultra-faint dwarf galaxies are reliable and can be meaningfully
compared with halo stars to investigate any potential chemical
differences.

\section{Abundances}\label{ab_pa}

\begin{figure*}
 \begin{center}
  \includegraphics[clip=true,width=18cm,bbllx=35, bblly=230, bburx=520,
   bbury=747]{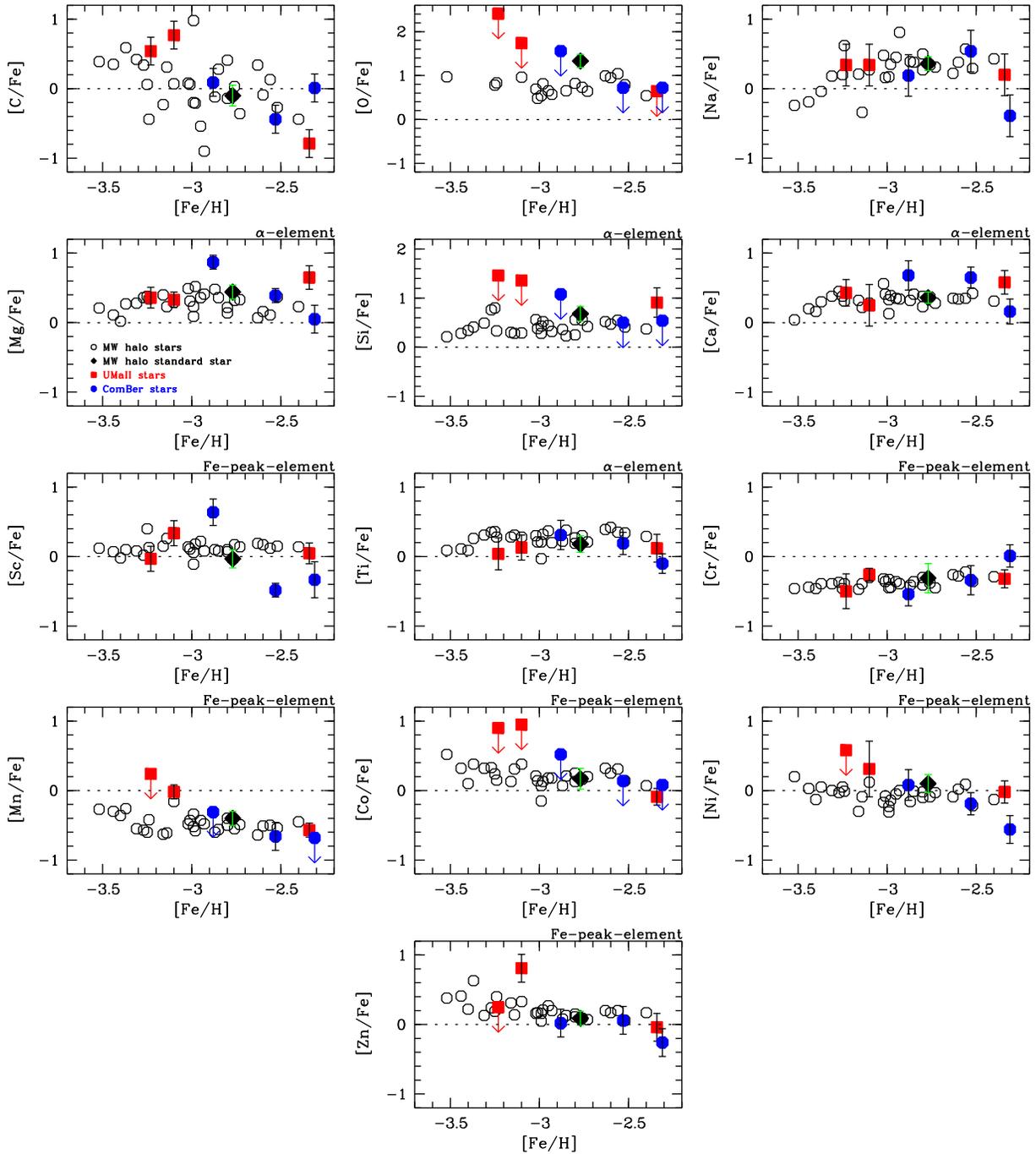} \figcaption{
     \label{cayrel_abundances_light} Abundance ratios ([X/Fe]) as a
     function of metallicity ([Fe/H]) for light and iron-peak elements
     in comparison with those of \citet{cayrel2004}. The y-axes of
     each panel have the same scale except for O and Si.  See
     \S\ref{sec:lightelements} for the discussion. \textit{Red
     squares} indicate UMa\,II stars, \textit{blue circles} show ComBer
     stars, \textit{open black circles} are the
     \citeauthor{cayrel2004} halo sample, and HD~122563, our Milky Way
     halo ``standard'' star, is shown by a black diamond. }
 \end{center}
\end{figure*}

\begin{figure*}
 \begin{center}
  \includegraphics[clip=true,width=16cm,bbllx=35, bblly=333, bburx=335,
   bbury=628]{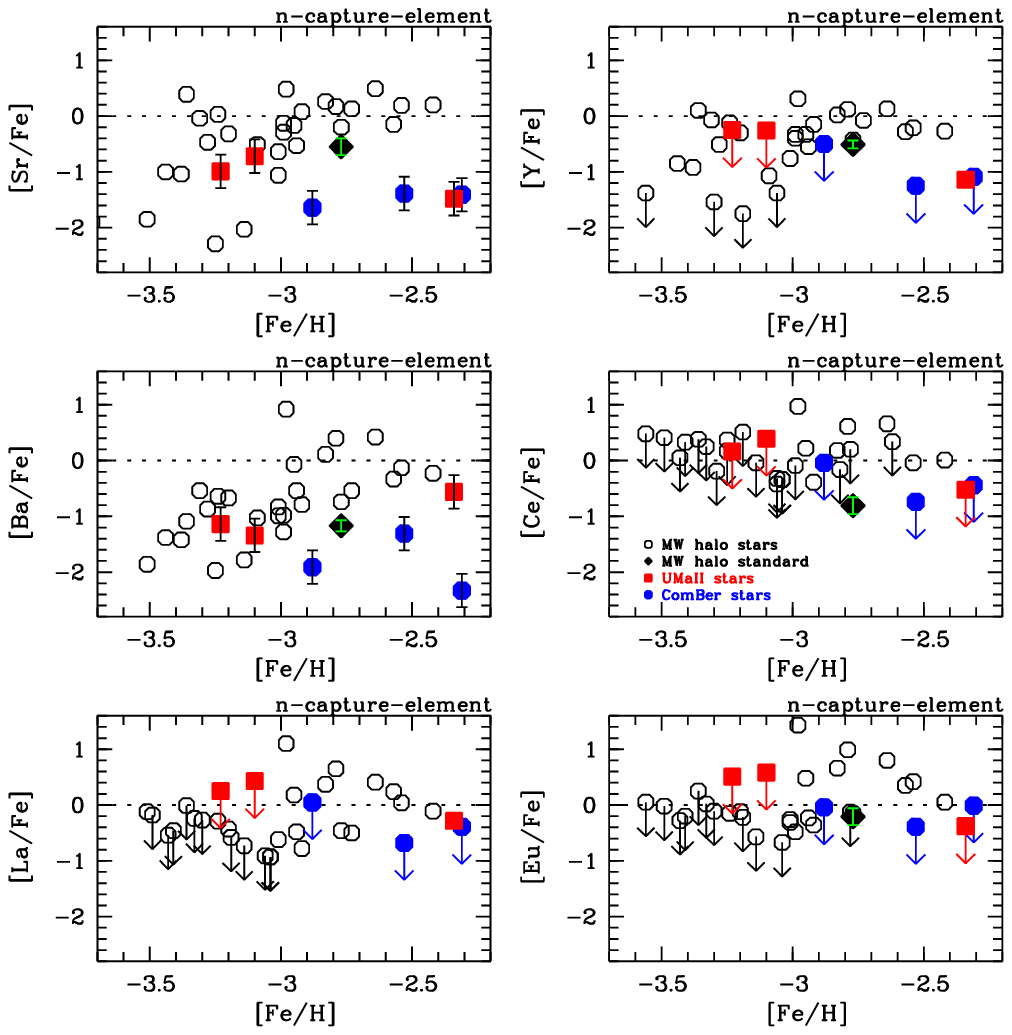} \figcaption{
   \label{cayrel_abundances_heavy} Abundance ratios ([X/Fe]) as a
   function of metallicity ([Fe/H]) for
   neutron-capture elements in comparison with those of
   \citet{cayrel2004}. 
   See \S\ref{nc} for the discussion. \textit{Red squares} indicate UMa\,II stars,
   \textit{blue circles} show ComBer stars, \textit{open black circles}
   are the \citeauthor{cayrel2004} halo sample, and HD~122563, our
   Milky Way halo ``standard'' star, is shown by a black diamond. Note that in
     the bottom plot, a Cayrel et al. star has an upper limit at the
     position of the Eu abundance of HD~122563.}
 \end{center}
\end{figure*}

In this section, we discuss the detailed abundance measurements for
each star in our sample and compare our results with MW halo stars of
similar Fe abundances.  The metal-poor stars in the MW halo are the
only known population with metallicities as low as what is found in
the ultra-faint dwarf galaxies. A useful comparison between the
abundance patterns of the two samples requires that the data sets
cover as similar a range in Fe abundance as possible.  Among the highest
quality data sets of metal-deficient MW halo giants available is the sample
of high $S/N$, high-resolution VLT/UVES spectra of 32 stars analyzed
by \citet{cayrel2004} (C to Zn) and \citet{francois07}
(neutron-capture elements).  A comparison of individual abundance
measurements in the ultra-faint dwarf galaxy stars in our study and
the \citeauthor{cayrel2004} and \citeauthor{francois07} MW halo stars
is presented in Figures~\ref{cayrel_abundances_light} and
\ref{cayrel_abundances_heavy}.

We begin each subsection below with a brief summary of the various
nucleosynthesis processes that produce the element or elements in
question.  \citet{woosley_weaver_1995} give an extensive description
of the production pathways for the interested reader.  A short summary
can also be found in \citet{cayrel2004}.

\subsection{Carbon}

Carbon is produced in the triple-$\alpha$ process during helium
burning on the red giant branch. Through dredge-up processes carbon is
transported to the surface where it can be lost to the ISM if strong
stellar winds are present.  Carbon is also expelled in supernova
explosions. The levels of C are driven by the explosion energy or the
amount of stellar rotation, and also depend on the mass of the
progenitor.  Measured C abundances in the most metal-poor stars thus
provide important fossil information on the various previous (early)
enrichment events and the nature of the first stars.  Rotating,
massive Population\,III stars \citep{meynet2005} may, for example,
have been significant producers of the first enrichments in CNO
elements.

\begin{figure}
  \begin{center}
    \includegraphics[clip=true,width=10cm,bbllx=47, bblly=140,
    bburx=525, bbury=746]{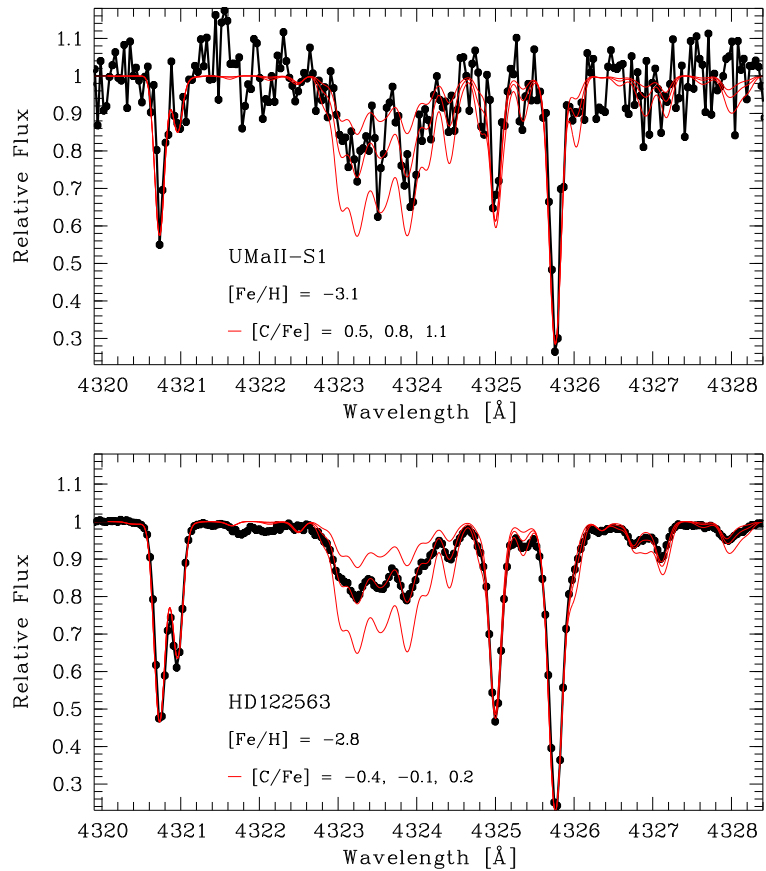} \figcaption{\label{ch4300} Spectral
      region around the CH feature at 4323\,{\AA} for UMa\,II-S1
      (upper panel) and HD122563 (lower panel). The observed spectrum
      is shown (\textit{thick line plus symbols}). Synthetic spectra with three
      different C abundances are shown in red. In addition to the
      Draco star Dra119 \citep{fulbright_rich}, these are the first
      carbon abundances measured in a dwarf galaxy.}
  \end{center}
 \end{figure}

We measure the C abundances of the dwarf galaxy stars from the G-band
head ($\sim4313$\,{\AA}) and the CH band at $4323$\,{\AA}. Except for
one measurement in Dra~119 \citep{fulbright_rich} no other stellar C
abundances are available in any of the dSphs or other dwarf galaxies.
To determine the C abundance, we compute synthetic spectra with
different C abundances and find the best fit with the observed
spectrum.  An example of the CH\,$\lambda4323$ synthesis is shown in
Figure~\ref{ch4300}.  The values we obtain for the two CH bands
generally agree with each other within $\sim0.2$\,dex, and we adopt
the average of the two measurements as our final C abundance.  The
dominant source of uncertainty for abundances derived from molecular
features is the continuum placement, especially in lower $S/N$
spectra. We thus assign a random uncertainty of
$\sigma\mbox{[C/Fe]}=0.15$. To check the validity of our abundance
scale we also determined the C abundance of HD~122563 (lower
panel of Figure~\ref{ch4300}). We derive an abundance of
$\mbox{[C/Fe]}=-0.1\pm0.1$ for this star. Bearing in mind that we
adopted somewhat different stellar parameters, and that abundances
from molecular features are quite temperature sensitive, this result is in
very good agreement with the value of
\mbox{$\mbox{[C/Fe]}=-0.35\pm0.2$} determined by \citet{aoki_studiesIV}.

We note that it is not possible to determine a $^{12}$C/$^{13}$C ratio
in our stars. The bottleneck for such measurements is the availability
of $^{13}$C lines. They are located at 4217.6 and 4225.2\,{\AA} where
the $S/N$ not sufficient to determine meaningful values or limits of
these usually very weak features (even for the stars that have the
highest C abundances in our sample).

\begin{figure*}
\begin{center}
 \includegraphics[clip=true, width=10.5cm, bbllx=54, bblly=80, bburx=440,
  bbury=715]{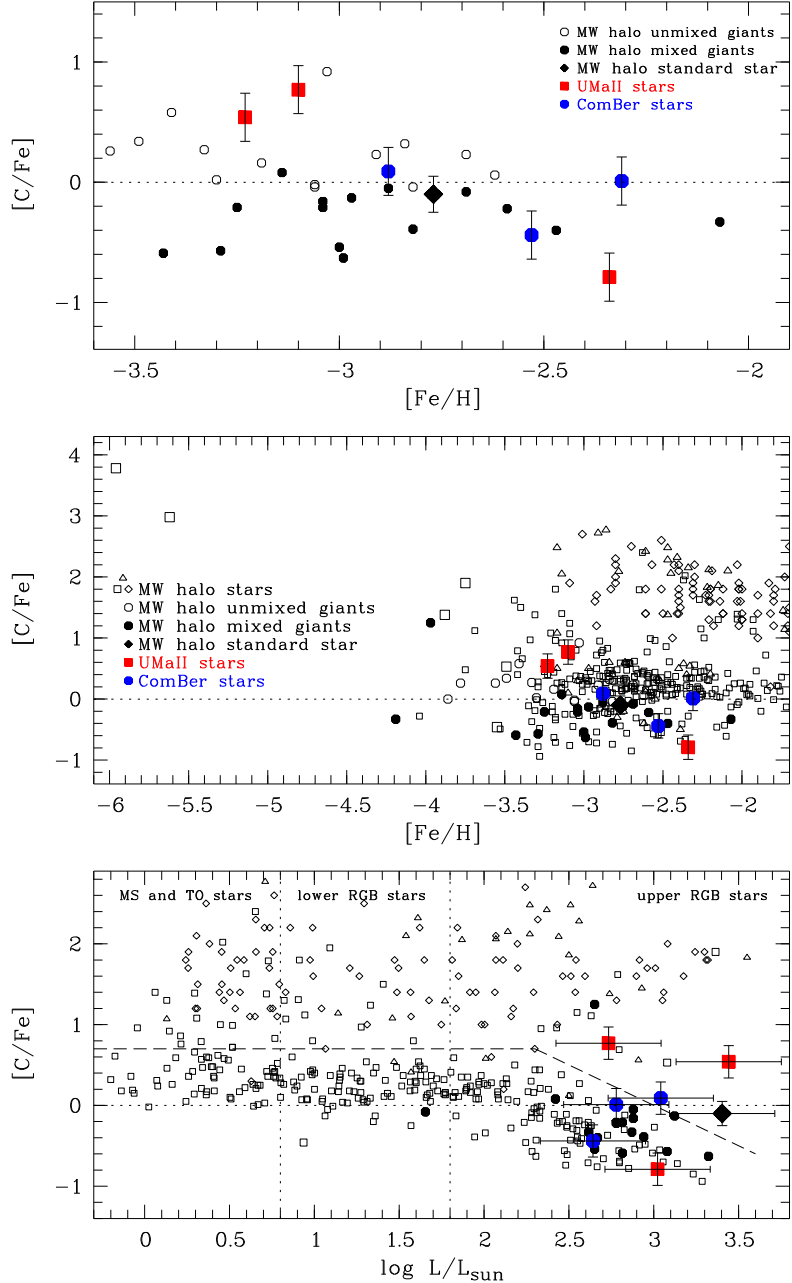} \figcaption{ \label{cfe} 
[C/Fe] abundance ratios as a function of
[Fe/H] (top and middle panels) and luminosity (bottom panel).   In all
three panels, blue circles indicate ComBer stars, red squares show
UMa\,II objects, and the black diamond is our MW halo standard HD~122563.
The top panel compares our C abundances to  MW halo giants
\citep{spite2006, aoki_cemp_2006} to assess the level of atmospheric
mixing which can affect the carbon abundances.  Middle panel:  The 
same comparison is shown with an expanded comparison sample  (squares:
\citealt{aoki_mg, cohen, heresII,collet06, frebel_he1300,
he1327_uves, lai2008}; triangles: \citealt{lucatello2006, aoki05}).    Bottom
panel: [C/Fe] ratios as a function of luminosity.  Indicated are   
different luminosity bins for each evolutionary stage. The  definition
of C-richness from \citet{aoki_cemp_2006} is also shown
(dashed line).   All three members of ComBer and the most metal-rich
UMa\,II star (UMa\,II-S3) are classified as carbon-normal, while the two more
metal-poor members of UMa\,II are carbon-rich.}
\end{center}
\end{figure*}

As seen in Figure~\ref{iso}, all our targets are on the upper red
giant branch (RGB), since less luminous stars in these galaxies are
too faint to observe at high resolution.  However, their somewhat
evolved nature may lead to concerns regarding the potential for
altered surface abundances. Such modifications (intrinsically through
dredge-up of nucleosynthesis products or extrinsically through binary
mass transfer), if not sufficiently quantified, could lead to
incorrect interpretations of the chemical nature of the gas from which
these stars formed. We note that in halo field stars, it can be
assumed that the abundances of elements other than C and N are not
affected by early signs of mixing (e.g., \citealt{spite2006}).
In the absence of a measured $^{12}$C/$^{13}$C ratio (which is often
used in infer details on the degree of atmospheric mixing and stellar
evolutionary effects) in our stars, we turn to the luminosities of our
targets to gain information on these effects. In Figure~\ref{cfe} (top
panel) we show the C abundances of our stars compared with
mixed\footnote{The separation of mixed and unmixed giants is based on
  a cut in N abundances at $\mbox{[N/Fe]=0.58}$, with the mixed stars
  having higher N abundances \citet{spite2006}.}  (solid circles) and
unmixed (open circles) metal-poor giants \citep[][also analyzed by
\citealt{spite2006}]{cayrel2004}. Their mixed giants are typical halo
stars that have ascended the giant branch far enough that early signs
of mixing can be observed.  As a star moves up the giant branch,
carbon decreases as a consequence of the CN cycling that converts C to
N.  This processing results in a drop of the C abundances with
increasing stellar luminosity \citep{gratton2000}.  Depletion values
can be as high as $\sim-0.7$\,dex for cool upper RGB stars such as our
targets.  The middle panel of Figure~\ref{cfe} shows the dwarf galaxy
C abundances compared with various halo stars from the literature to
allow the reader to appreciate where the stars in our dwarf galaxies
are located with respect to the present range of halo data.  Our C
abundances agree well with those of the mixed halo giants, except for
the two most metal-poor stars in our sample.  Those stars have much
higher C abundances, similar to what is found for unmixed
stars. However, all our stars have lower effective temperatures (i.e.,
the stars are further evolved) than those of Cayrel et al. (2004)
which suggests that all of them must have undergone some degree of
mixing, even the two most metal-poor stars.  This ``discrepancy'' thus
indicates that they appear to be C-enhanced beyond what is expected
from canonical stellar evolution (i.e., the level of the mixed
stars). This would mean that the stars must have been born from
material that was overabundant in carbon.  We note that the Spite et
al.~stars are not C-enriched, i.e. they formed from gas that was not
enriched in carbon beyond the general level provided by the chemical
evolution at that time.  The common definition of C-rich metal-poor
stars is $\mbox{[C/Fe]}>1.0$ (e.g., \citealt{ARAA}), assuming that the
currently observed C abundance reflects the abundance of the birth
material. Our stars do not quite reach this level, but as we discuss
below, for upper red giant branch this
``one-size-fits-all''-definition may not be appropriate because mixing
processes change the surface C abundances we observe today.

In the bottom panel of Figure~\ref{cfe}, we plot the [C/Fe] abundances
as a function of luminosity. Taking the luminosity dependent
C-decrease into account, \citet{aoki_cemp_2006} suggested redefining
the classification of C-rich stars to\\

\noindent

$\mbox{[C/Fe]} \ge +0.7$ \mbox{for stars with} $\log (L/L_{\odot} ) \le 2.3$ \\

\noindent

and\\

\noindent

$\mbox{[C/Fe]} \ge +3.0 - \log(L/L_{\odot} )$ \mbox{when}
  $\log(L/L_{\odot}) > 2.3$.\\

\noindent
This boundary is indicated in Figure~\ref{cfe} (dashed line). As can
be seen, all three members of ComBer are not enriched in C; in fact,
they all must have had very similar C abundances at birth that by
now have decreased to the observed level because of their large
luminosities. The two most metal-poor stars (both in UMa\,II), on the
other hand, can be classified as C-rich. They contain larger C
abundances than what is expected for stars at larger luminosities
(i.e. low effective temperatures) in which some mixing has been taken
place.  The third UMa\,II object, however, is rather like the ComBer
stars, i.e., it shows a decrease in C consistent with its luminosity
and must have been born from material that was not especially enriched
in carbon.  This strongly suggests a large spread of C abundances
in UMa\,II, which could point to either different production sites and
timescales (the C-normal star is more metal-rich than the C-rich
stars) or varying degrees of mixing in this galaxy.  We see no obvious
correlation between Fe abundance and the galactocentric distance of
the stars.  This indicates that an ad hoc assumption of incomplete
mixing in the ISM may not explain the very large Fe and C
spread. However, with the small existing sample of stars and no
knowledge about the actual physical boundaries and dynamical history
of these systems we can only speculate whether the present location of
our targets would reveal anything at all about mixing processes.  More
stars in both of these systems are clearly needed to provide more
insight into this issue.  Nevertheless, the fact that two of our six
stars are C-rich is very interesting in itself. As has been known for
quite a while, a large fraction of metal-poor halo stars are enriched
in C, with numbers ranging from $\sim15\%$ \citep{frebel_bmps, cohen}
to $\sim25\%$ \citep{marsteller}. Below $\mbox{[Fe/H]}<-3.0$, these
numbers are found to increase in all samples, although those results
are generally plagued by small-number statistics. Hence, finding two
C-rich stars in a dwarf galaxy, which are also the two most metal-poor
stars in our sample and among the most metal-poor ones in the entire
sample of \citet{kirby08}, is suggestive of a high fraction of C-rich
stars in dwarf galaxies as well.  Although the statistical
significance of this result is low, it may indicate that C generally
played an important role in the formation and evolutionary process at
early times irrespective of the host galaxy system.

\subsection{Elements with $Z\le30$}
\label{sec:lightelements}

Light elements are produced during stellar evolution or directly in
supernovae and then expelled during the explosions (e.g.,
\citealt{woosley_weaver_1995,nomoto97}). 

\subsubsection{Sodium}
Na is produced during carbon burning and through the Ne-Na cycle
during H burning \citep{woosley_weaver_1995}. It has thus been
suggested that Na correlates with Ni since the Ni production depends
on the neutron excess provided by $^{23}$Na during the supernova
explosion that drives the $^{58}$Ni abundances (see also
\citealt{venn04}). This hypothesis would explain why both elements are
observed to have similar abundances in stars.

The \ion{Na}{1}\,D resonance lines at $\sim$5890\,{\AA} are used to
determine the Na abundances. The resonance lines are very sensitive to
non-LTE effects. We note that, in principle, all [Na/Fe] abundances
shown in Figure~\ref{cayrel_abundances_light} should be decreased by
several tenths of dex (e.g., \citealt{na_nlte_baum}) to account for
non-LTE effects. For ease of discussion, however, we simply compare
our LTE Na abundances with the uncorrected (i.e., LTE) abundances of
\citet{cayrel2004}.  The (LTE) agreement is generally quite good,
although our highest metallicity star deviates from the bulk of the
halo and our other dwarf galaxy abundances by almost one dex. Such low
Na abundance is very unusual, and no other stars are known with
similarly low Na values.

\begin{figure}
 \begin{center}
   \includegraphics[clip=true,width=8cm,bbllx=30, bblly=410,
   bburx=452, bbury=714]{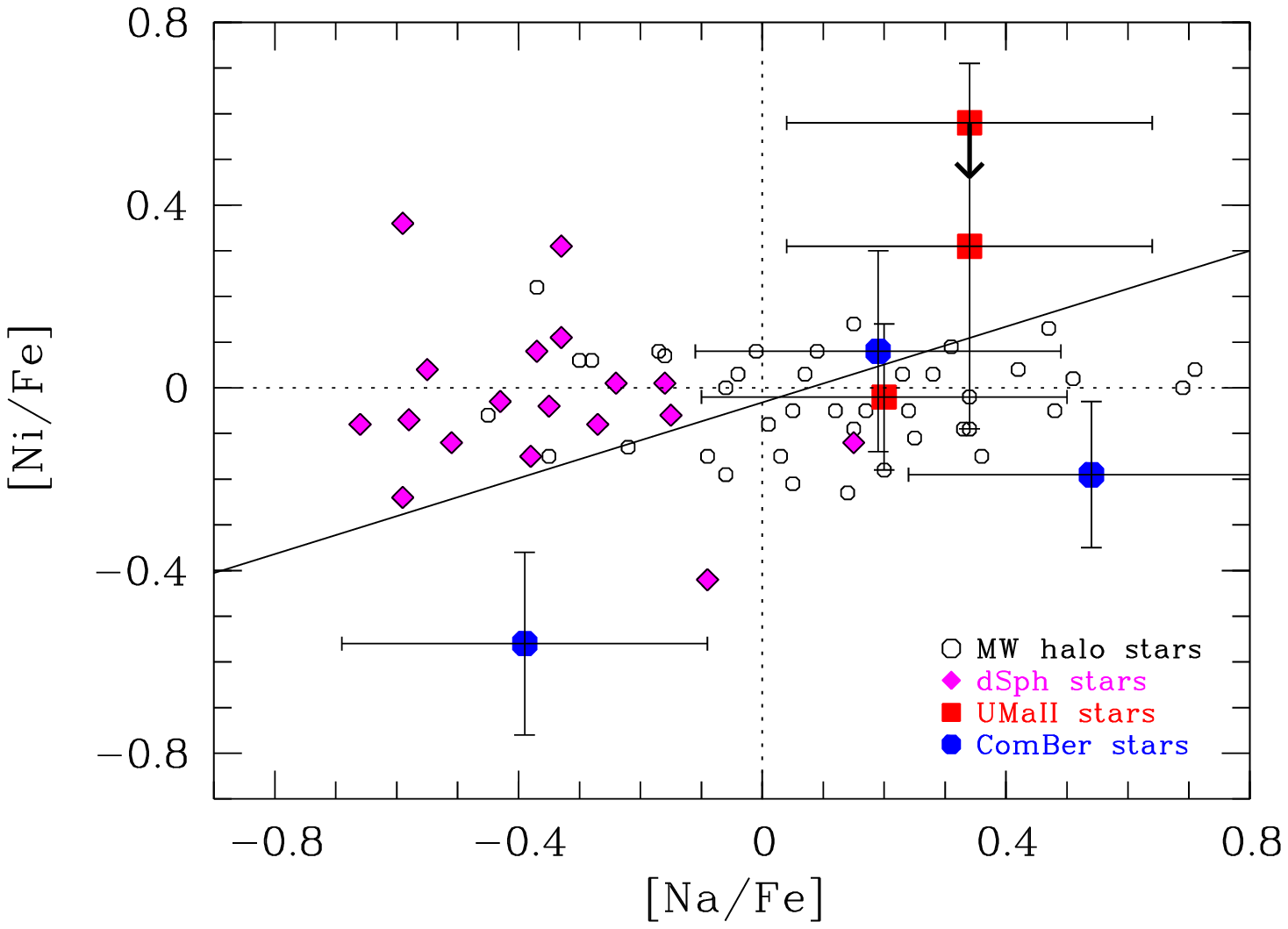} \figcaption{ \label{nina}[Ni/Fe]
     ratios for the program stars as function of [Na/Fe] in comparison
     with other objects from the literature (black circles:
     \citealt{cayrel2004, lai2008}; pink diamonds: \citealt{venn04}). Blue
     circles indicate ComBer stars, whereas red squares indicate UMa\,II
     stars. The relation found by \citet{nissen_schuster} is also
     shown (solid line). The two UMa\,II stars that deviate most from
     the Nissen \& Schuster relation are also the only two carbon-rich
     stars in the sample.}
 \end{center}
\end{figure}

\citet{nissen_schuster} analyzed a disk star sample with somewhat
deficient [Na/Fe] and [Ni/Fe] abundances. Figure~\ref{nina}
illustrates the correlations they found and shows that ComBer-S1
extends this relationship to lower Na and Ni values by $\sim0.5$\,dex.
Our other stars do not deviate significantly from the suggested
correlation.  There is a large scatter among the halo stars of
\citet{cayrel2004} (black symbols in the Figure), and it is difficult
to evaluate whether they generally follow the \citet{nissen_schuster}
trend.  It should also be said that the \citeauthor{cayrel2004} stars
are halo stars and have much lower Fe abundances than the stars
originally considered for this relationship.  Nevertheless, they seem
to follow the trend better than the higher-metallicity dSph stars
collected by \citet{venn04}, which generally have higher Ni abundances
than predicted by the correlation.  This may indicate different
nucleosynthetic origins for the stars in the brighter dSphs and is
discussed further below. 

\subsubsection{$\alpha$-Elements}
The $\alpha$-elements (Mg, Ca, Si, Ti) are built from multiples of He
nuclei since they are produced through $\alpha$-captures during
various burning stages of stellar evolution (carbon burning, neon
burning, complete and incomplete Si burning) and then dispersed during
the explosions of core-collapse supernovae.  Although Ti ($Z=22$) is
not a true $\alpha$-element, in metal-poor stars the dominant isotope
is $^{48}$Ti, which behaves like an $\alpha$-element.

\begin{figure*}
 \begin{center}
   \includegraphics[clip=true,width=12cm,bbllx=47, bblly=110,
   bburx=540, bbury=300]{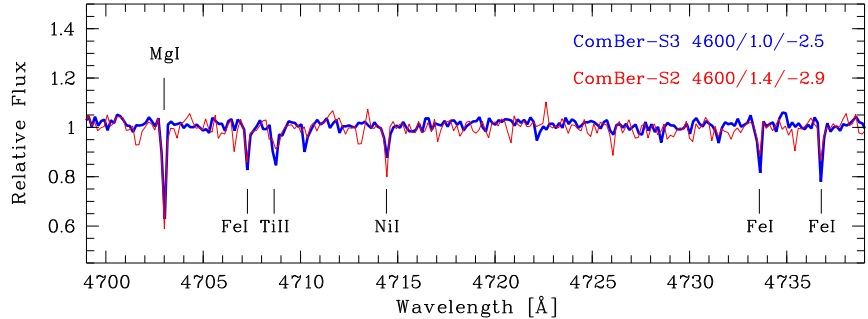} \figcaption{\label{mg_region}
     Spectral region around 4700\,{\AA} in ComBer-S2 in comparison to
     ComBer-S3. The stars have similar stellar parameters but
     different Fe abundances, which is reflected in the different Fe line
     strengths, except for the Mg line at 4703\,{\AA} which is
     stronger in ComBer-S2. Several species are labeled, and the
     atmospheric parameters of both stars are given in the legend.}
 \end{center}
\end{figure*}

Several \ion{Mg}{1} lines across the spectrum were employed to derive
the Mg abundance. In some cases, the \ion{Mg}{1}\,b triplet lines at
$\sim$5170\,{\AA} were very strong and beyond the linear part of the
curve of growth. Nevertheless, the abundances of those lines generally
agreed with those of the other Mg lines. Four of our six stars have Mg
abundances of $\mbox{[Mg/Fe]}\sim0.4$, in good agreement (see
Figure~\ref{cayrel_abundances_light}) with the general trend in MW
halo stars (e.g., \citealt{cayrel2004, heresII, lai2008}). Two stars,
however, are overabundant in this element: ComBer-S2 has
$\mbox{[Mg/Fe]}\sim1.0$, and UMa\,II-S3 has
$\mbox{[Mg/Fe]}\sim0.7$. Figure~\ref{mg_region} shows the spectral
region around the a \ion{Mg}{1} line at 4703\,{\AA} of ComBer-S2 in
comparison with ComBer-S3, one of the stars with a lower, MW halo-like
Mg abundance. The stars have the same effective temperature and
similar surface gravities, but different Fe abundances. This leads to
the different line strengths for all metals, which can be seen in the
figure. The Mg lines, however, have roughly the same strength,
illustrating the Mg-rich nature of ComBer-S2. There are a few cases
known where metal-poor MW halo stars have large Mg overabundances of
up to $\mbox{[Mg/Fe]}\sim2.0$ (e.g., \citealt{aoki_mg,
  HE1327_Nature}). Some of these stars are very C-rich as well, but
that is not the case for ComBer-S2. Si may also be enhanced in such
stars, but unfortunately our spectra do not cover the strong Si line
at 3905\,{\AA}. Hence, we are only able to derive upper limits from
the much weaker 5684\,{\AA} line. The limit for the most Mg-rich star
(ComBer-S2) rules out a Si abundance of $\mbox{[Si/Fe]}\gtrsim1.0$.

\begin{figure*}
 \begin{center}
  \includegraphics[clip=true, width=10cm,bbllx=50, bblly=78,
   bburx=443, bbury=742]{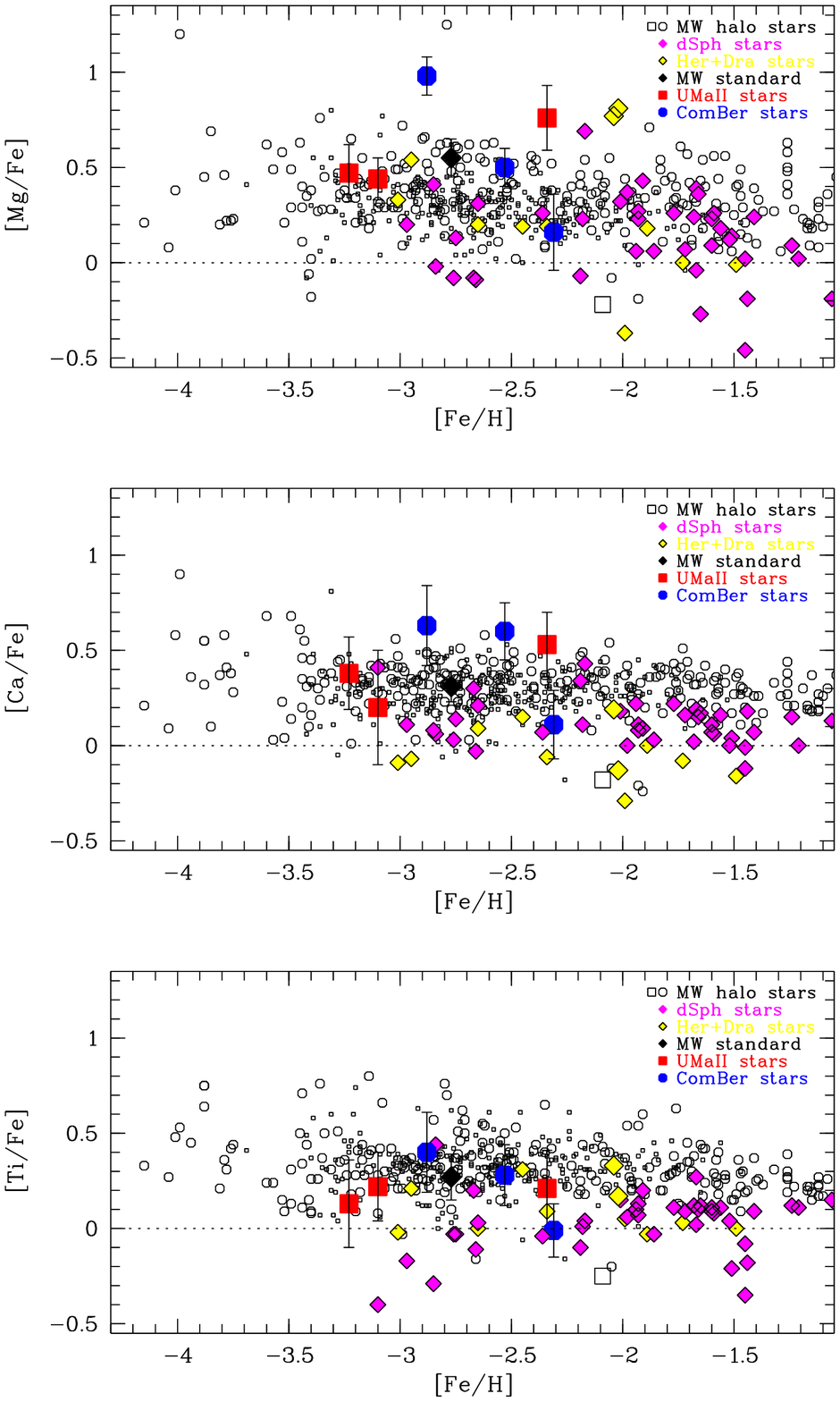} \figcaption{
     \label{mgcati}[Mg/Fe] (top panel), [Ca/Fe] (middle panel) and
     [Ti/Fe] (bottom panel) abundance ratios as a function of
     [Fe/H]. Generally, the ultra-faint dwarf galaxy abundances
     (\textit{red squares}: UMa\,II stars; \textit{blue circles}:
     ComBer stars) agree with the metal-poor halo abundances (black
     squares and circles), in contrast to those of the more luminous
     dSphs (\textit{pink diamonds}; \citealt{shetrone01, shetrone03,
     sadakane04, aoki09}). We also find two stars that are Mg-rich,
     similar to a few known halo stars.  Halo data are taken from
     \citet{lai2008}, \citet{francois07}, \citet{heresII}, and
     \citet{venn04}. The metal-poor halo standard star HD122563 is
     marked with a black diamond. The big yellow diamonds at
     $\mbox{[Fe/H]}\sim-2.0$ are two stars in the ultra-faint dwarf
     galaxy Hercules \citep{koch_her}. The small yellow diamonds refer
     to the Draco data by \citet{cohen09} and Draco D119 at
     $\mbox{[Fe/H]}\sim-2.0$ \citep{fulbright_rich}, which show
     somewhat similar Mg and Ti abundances to our targets. The open
     black square indicates the $\alpha$-poor, neutron-capture-poor
     star BD 80 245 \citep{ivans_alphapoor}. }
 \end{center}
\end{figure*}

The Ca abundances are shown in Figure~\ref{cayrel_abundances_light},
derived from several lines of \ion{Ca}{1}. Similar to Mg, the Ca
abundances agree very well with the MW halo trend of
$\mbox{[Ca/Fe]}\sim0.4$. The Ti abundances are based on numerous lines
across the spectrum. Generally, \ion{Ti}{1} values agree within
$0.2$\,dex to those of \ion{Ti}{2}, which in turn are in good
agreement with the \citet{cayrel2004} halo pattern
(Figure~\ref{cayrel_abundances_light}).
Figure~\ref{mgcati} shows a more detailed comparison of our
$\alpha$-element results with those of a large set of literature halo
stars. While there are some outliers among our more metal-poor stars
that deviate from the halo data by having slightly \textit{higher}
abundances, the general trend of the more luminous dSphs to have
[$\alpha$/Fe] ratios \textit{below} the halo data is not followed by
UMa\,II and ComBer.

\subsubsection{Iron-peak Elements}
In the early universe, the iron-peak elements (Sc to Zn; $23\le Z
\le30$) are exclusively synthesized during Type~II supernova (SN\,II)
explosions by explosive oxygen and neon burning, and complete and
incomplete explosive Si burning.  Only at later times, once the stars
less massive than those exploding as SNe\,II reach the end of their
life time, do SNe\,Ia became the dominant contributor to the total
iron inventory. The onset of SNe\,Ia in the chemical evolution of the
MW halo is clearly observed in the [$\alpha$/Fe] vs. [Fe/H] plane by
means of a down-turn of the  $\mbox{[$\alpha$/Fe]}\sim0.4$ plateau at
metallicities above $\mbox{[Fe/H]}\sim-1.5$ (e.g.,
\citealt{McWilliametal, ryan96}).

We measured the Fe-peak elements Sc, Cr, Mn, Ni, Fe and Zn in our
dwarf galaxy stars. Overall, there is good agreement between our
abundances and those in the halo.  Sc and Mn abundances were
determined from several lines. Hyper-fine structure was taken into
account (using Kurucz line lists) and the abundances of the lines were
derived from spectral synthesis. The Sc abundances in UMa\,II agree
with those of the halo stars, but there is significant scatter found
in ComBer (up to $\sim1$\,dex). Mn, Ni and Zn abundances were
determined, respectively, in 3, 5 and 5 of the 6 program stars, and
upper limits were derived for the remaining stars. Our measured Mn,
Cr, Ni, and Zn abundances generally follow the halo trend. The only
exception is the highest metallicity star (ComBer-S1), which has a
rather low Mn upper limit and a low, subsolar Ni abundance that can be
explained by the low Na abundance. The most metal-poor star,
UMa\,II-S2, has an unusually high Zn abundance that could indicate a
very high explosion energy for the supernovae responsible for its
abundance pattern \citep{UmedaNomoto:2002}.

\subsubsection{Upper Limits}

Upper limits were determined for O, Al, Si, V, Co, and Cu. The limits
are generally tighter at higher Fe values where the searched-for
lines are expected to be stronger. For Co, the limits indicate no
enhancement with respect to the halo material among our higher
metallicity stars, and perhaps a small deficit. We note for
completeness that all our stars are too evolved to show any detectable
Li in their spectra. Because of the increased thickness of the
convection zone as the stars ascend the giant branch, the Li becomes
diluted and destroyed as it mixes into deeper, hotter layers.

\subsection{Neutron-Capture Elements}\label{nc}

Neutron-capture elements (Sr to U, $38\le Z \le92$) can originate from
a variety of nucleosynthetic processes. It is thus not easy to
disentangle the different sources and arrive at meaningful
conclusions. The two major pathways for the production of these
elements are the rapid (r) process thought to occur in supernova
explosions and the slow (s) process thought to occur in AGB stars
during stellar evolution. A detailed review of the importance of
neutron-capture elements and their abundance in the MW halo can be
found in \citet{sneden_araa}.

\subsubsection{Strontium and Barium}

\begin{figure*}
 \begin{center}
   \includegraphics[clip=true,width=10cm,bbllx=50, bblly=278,
   bburx=443, bbury=742]{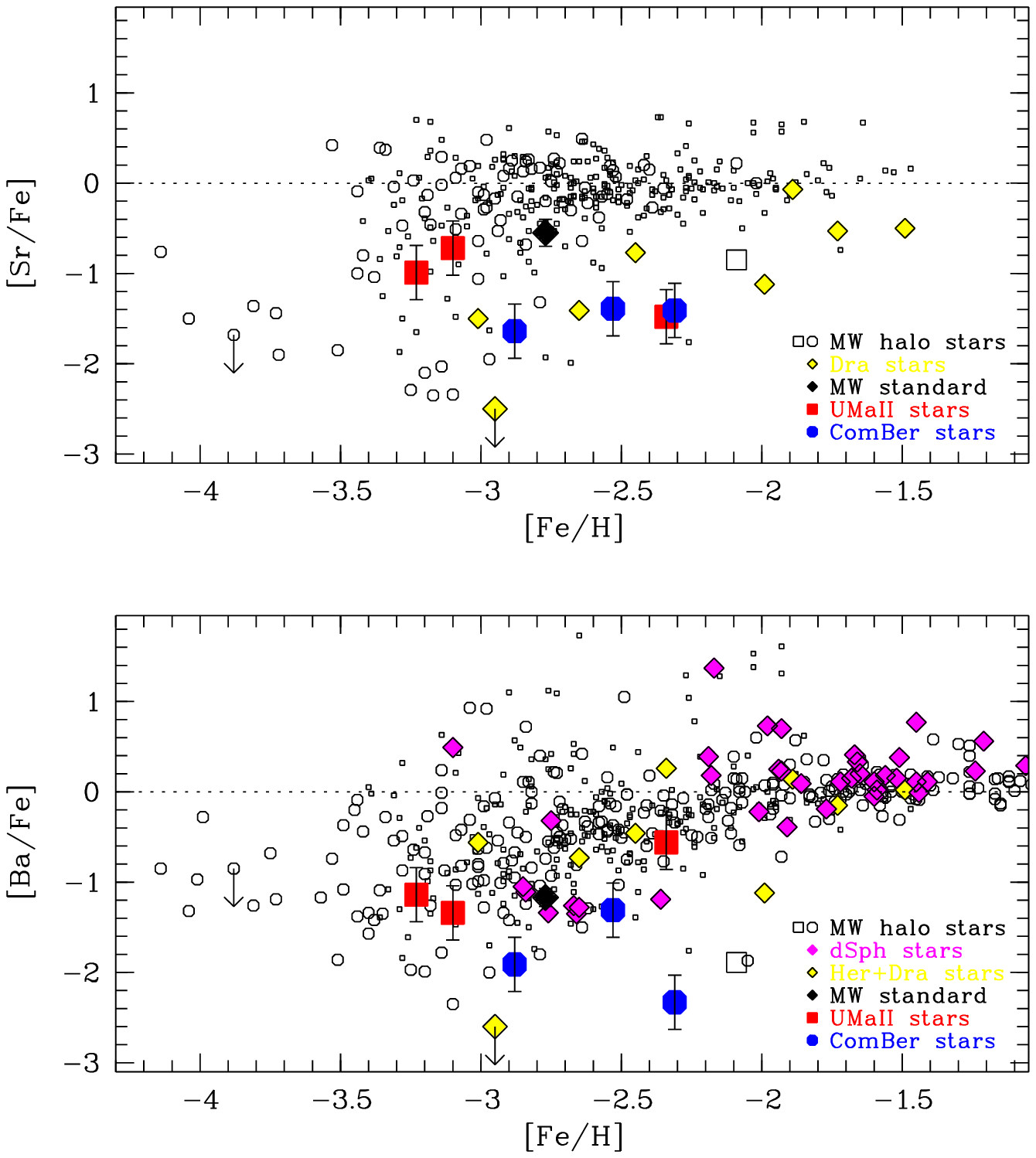} \figcaption{ \label{srbafe} [Sr/Fe]
     (top panel) and [Ba/Fe] (bottom panel) abundance ratios as a
     function of [Fe/H]. Except for the upper limit in Dra119
     (\textit{yellow diamond}; \citealt{fulbright_rich}), these are
     the first Sr measurements in dwarf galaxies.  All the ultra-faint
     dwarf galaxy [Sr/Fe] and [Ba/Fe] abundances (\textit{red squares}: UMa\,II
     stars; \textit{blue circles}: ComBer stars) are very low, at the
     lower end of the distribution of metal-poor halo stars (black squares and
     circles). Symbols are the same as in Fig.~\ref{mgcati}. }
 \end{center}
\end{figure*}

We measure the neutron-capture elements Sr and Ba in all of our stars.
We find that both the [Sr/Fe] and [Ba/Fe] ratios are extremely
depleted compared with the overall MW halo pattern. Low Ba abundances
have also been found by \citet{shetrone01} for a few stars in the more
luminous dSphs.  It was suspected that Sr might be similarly depleted,
but no Sr measurements exist because of the limited spectral coverage
of those observations.  We provide the first Sr measurements in any
dwarf galaxy\footnote{During the completion of this paper,
  \citet{cohen09} also obtained Sr abundances for their sample of
  Draco stars.}, and confirm that the Sr values are indeed at a
similar level as Ba. This result is not surprising since the
nucleosynthetic origin of Ba and Sr is expected to be the same. In
ComBer, Sr is almost constant at $\mbox{[Sr/Fe]}\sim-1.5$ over the
metallicity range $-2.9<\mbox{[Fe/H]}<-2.3$. These values are very
low, more than 1\,dex below the bulk of the halo stars at the same Fe
abundances (although as can be seen in
Figures~\ref{cayrel_abundances_heavy} and \ref{srbafe}, there are a
few halo stars with similarly low levels). There is, however, a
significant scatter of several dex (up to $\sim3$\,dex) among halo
stars at the lowest metallicities that is not well understood.
\citet{francois07} find that the neutron-capture abundances generally
decrease at the lowest Fe abundances, i.e., below
$\mbox{[Fe/H]}\sim-3.0$. Furthermore, most of the stars with
$\mbox{[Fe/H]}\lesssim-3.5$ have very low neutron-capture abundances
(e.g., Sr and Ba; \citealt{Norrisetal:2001}), and this trend is also
found in two of the three halo stars known with $\mbox{[Fe/H]}<-4.0$
\citep{HE0107_ApJ, he0557}.

In contrast to the low and uniform Sr abundances in ComBer, UMa\,II
interestingly shows substantially higher [Sr/Fe] ratios in the two
stars with $\mbox{[Fe/H]}\sim-3.2$ ($\mbox{[Sr/Fe]}\sim-0.8$) than in
the star with $\mbox{[Fe/H]}\sim-2.3$ ($\mbox{[Sr/Fe]}\sim-1.3$,
similar to the ComBer stars).  The Sr values of the two EMP stars fit
well into the range seen in halo stars.  Generally, the scatter of Sr
abundances strongly increases with decreasing metallicity [Fe/H], with
more and more stars having [Sr/Fe] abundances much lower than the
solar value at low metallicities.  Compared with the dwarf galaxy
stars, HD~122563 is significantly Sr-enriched at
$\mbox{[Sr/Fe]}\sim-0.6$, although this is still deficient by a factor
of four relative to Fe compared to the Sun.

Even though all of our targets are deficient in Ba, we also observe is
a pronounced scatter in the [Ba/Fe] values. The constant trend of Sr
abundances in ComBer is not followed in Ba, with up to 1.5\,dex of
scatter (see Figure~\ref{srbafe}).  All three ComBer stars have Ba
abundances at or below the lower envelope of the halo stars; one star,
ComBer-S1, is well below the entire Cayrel et al. (2004) halo sample,
at $\mbox{[Ba/Fe]}=-2.33$. The Ba $\lambda4554$~\AA\ line in this
star is quite weak, as can be seen in Figure~\ref{ba4554_spec},
although the detection is significant at the 3~$\sigma$ level.
Conservatively, one could regard this measurement as an upper limit,
which would suggest an even more extreme underabundance of Ba.
UMa\,II is somewhat different from ComBer, with its two most
metal-poor stars also having [Ba/Fe] ratios towards the low end of the
halo distribution; the star at $\mbox{[Fe/H]}\sim-2.3$ has a much
higher, almost solar [Ba/Fe] ratio, although the abundances of this
star may have a somewhat different origin (see \S~\ref{rs}).

Incomplete mixing could explain the significant differences among
stars in each of our two dwarf galaxies in Ba, and to some extent also
in Sr.  What is telling, though, is that with minor exceptions,
\emph{all} of the neutron-capture elements in the ultra-faint dwarf
galaxies are at the same level as the lowest abundances found in the
halo.

\subsubsection{Upper limits}

For each of the stars we determined upper limits for Y, Zr, La, Ce,
Nd, Sm, and Eu (with the exception of UMa\,II-S3, in which we were
able to detect Y and La). They are listed in Tables~\ref{Tab:Eqw}, and
\ref{abundances_uma} and \ref{abundances_comber}. The Y, Ce and Eu
limits are compared with halo abundances and limits in
Figure~\ref{cayrel_abundances_heavy}. Generally, the limits indicate
deficiencies in neutron-capture elements relative to the halo
(particularly for our more metal-rich targets), consistent with the
low Sr and Ba values. [Y/Fe], [Ce/Fe] and [Eu/Fe] in our more
metal-rich stars are deficient by more than $\sim-0.5$ to
$-1$\,dex. Much higher $S/N$ data are needed to obtain more stringent
limits and to explore whether all the neutron-capture elements in the
ultra-faint dwarfs have depletion levels of $\sim-$2\,dex with respect
to the solar value.

\subsubsection{Origin of the Heavy Elements in the \\Ultra-Faint Dwarf
  Galaxies}\label{heavy}

We now use the observed neutron-capture abundances of our target stars to
infer information about the different nucleosynthetic processes that
played a role in the early history of the ultra-faint dwarf
galaxies. At low metallicity, a major distinction can be made
between the r- and s-process signatures, indicating early SN\,II
(pre-) enrichment or later mass transfer events, respectively.

Three of our stars have Fe abundances above the threshold value of
$\mbox{[Fe/H]}\sim-2.6$ at which the s-process sets in for halo stars
\citep{simmerer2004}, while the other three have lower metallicities.
In principle, this suggests that the s-process could be responsible
for the observed neutron-capture abundance patterns of the
higher-metallicity half of our sample. Indeed, one of our stars
(UMa\,II-S3, with $\mbox{[Fe/H]}\sim-2.3$) may have a neutron-capture
pattern consistent with an s-process signature, although its overall
neutron-capture abundances are very low (usually the s-rich metal-poor
stars exhibit [neutron-capture/Fe] values of $>0$). Since the
neutron-capture abundances are so low it is not entirely clear whether
the s-process of a previous generation of AGB stars could have
enriched the gas cloud with s-material (e.g., through mass loss) from
which our target formed, or if UMa\,II-S3 received this material from
a binary companion. UMa\,II-S3 does, however,  exhibits the typical radial
velocity variations indicating binarity (see \S~\ref{rs}).

At low metallicities, the r-process is a promising candidate for the
origin of the neutron-capture elements since it is associated with
massive SNe\,II that are expected to have been present at very early
times.  The low Fe abundances ($\mbox{[Fe/H]}<-2.6$) of our three most
metal-poor target stars thus indicate that the gas from which they
formed was probably enriched through the r-process by SNe\,II from the
previous generation of stars.  As for the two more metal-rich stars
(setting aside UMa\,II-S3 for the moment), their very low Sr and Ba
abundances may suggest that they too originated from gas enriched by
the r-process.

In recent years, it has been suggested that there are two components
of the r-process, which produce somewhat different neutron-capture
abundance distributions \citep{travaglio, aoki05, otsuki}. The weak
r-process is thought to produce mainly the lighter neutron-capture
elements ($Z<56$) and little or no heavier neutron-capture elements,
such as Ba. The so-called main r-process, on the other hand, produces
the full range of neutron-capture elements up to $Z = 92$. The weak
r-process has been suggested to occur in massive
($\gtrsim20$\,M$_{\odot}$) core-collapse SNe (e.g.,
\citealt{wanajo_ishimaru, izutani}), whereas the main component may
occur in supernovae with lower mass (8-10\,M$_{\odot}$) progenitors
\citep{qian_wasserburg03}.  The two different signatures are
principally observable in suitable stars\footnote{Objects that
  exhibit strong overabundances in neutron-capture elements, have low
  effective temperature, and are bright enough  for
  high-resolution, high $S/N$ spectra to be acquired.}, and, for
example, HD~122563 has been suspected of exhibiting a weak r-process
signature because the abundance of Ba and heavier elements are
depleted \citep{honda06}. From examining the [Ba/Sr] ratio,
which reflects the relative contributions of the two processes, clues
can be obtained as to the potential origin of the overall abundance
pattern. If the main r-process were at work, higher overall levels of
Sr and Ba would be expected (e.g., \citealt{honda07}).

\begin{figure}
 \begin{center}
   \includegraphics[clip=true,width=8.5cm,bbllx=20, bblly=400,
   bburx=450, bbury=640]{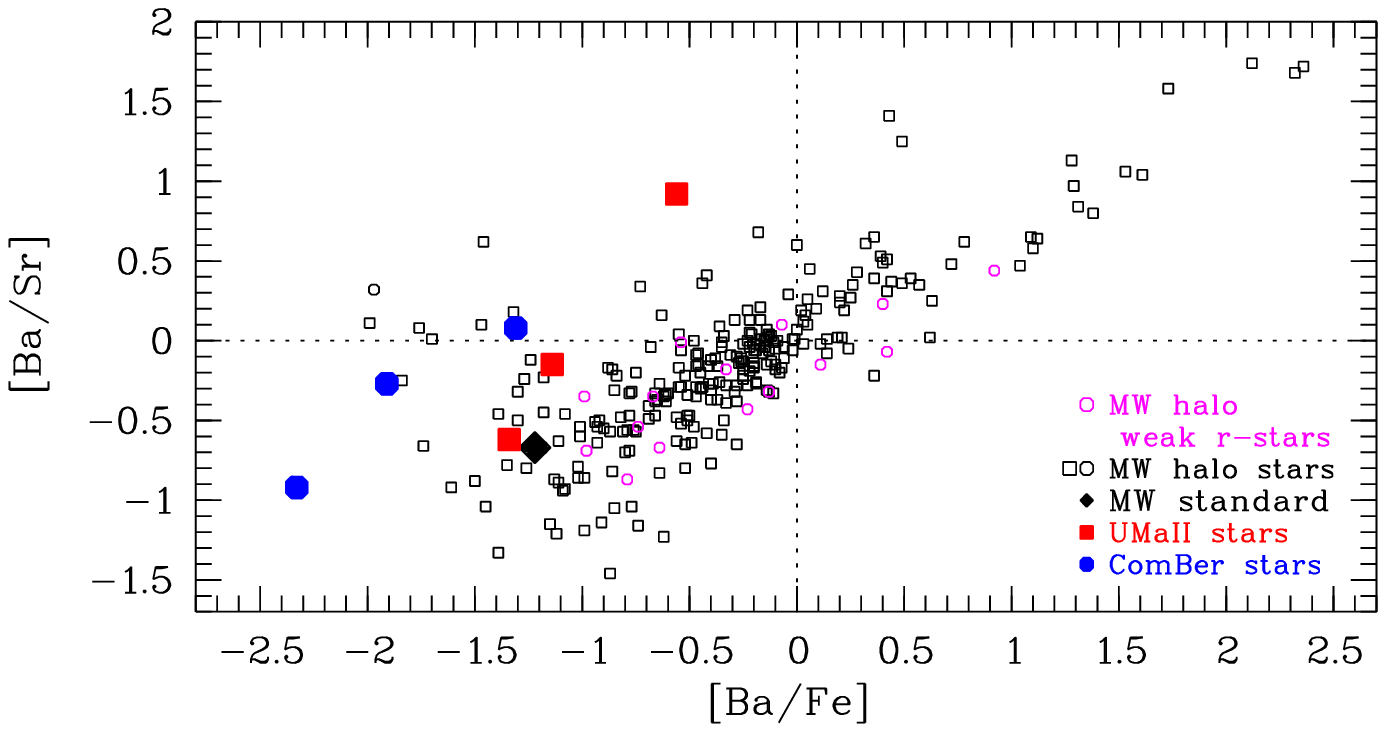} \figcaption{ \label{srba}[Ba/Sr]
     ratios for the program stars as a function of [Ba/Fe] in comparison
     with other objects (black symbols) from \citet{heresII, lai2008}
     and \citet{francois07}. Red squares indicate UMa\,II stars, blue
     circles show ComBer objects. The potentially weak r-process
     enriched HD122563 is marked with an open black diamond. The full
     pink circles are Francois et al. stars classified as weak
     r-process stars by \citet{izutani}. At least two of our stars
     have [Ba/Sr] values consistent with those of the Francois
     et al. weak-r halo stars and HD122563.}
 \end{center}
\end{figure}

The extremely low Ba abundances (and somewhat higher Sr levels) we
observe are thus suggestive of the weak r-process as the most likely
nucleosynthetic origin for the neutron-capture elements in UMa\,II and
ComBer.  In Figure~\ref{srba} we compare the [Ba/Sr] ratios of our
stars with the MW halo ratios.  We find that our targets have similar
abundance ratios to HD~122563, indicating that the explosions of
very massive stars might have provided the early chemical enrichment in
both UMa\,II and ComBer.  This would be in accord with the
halo-typical enhancements of $\alpha$-elements that also originate
from nucleosynthesis in massive stars (e.g.,
\citealt{McWilliametal}). The production of Fe is decoupled from that
of the neutron-capture elements (e.g., \citealt{sneden_araa}), so one would
still have to explain the spread in Fe found in our sample to arrive at
a global explanation for the chemical abundance patterns of these
systems.  Fe abundances around $\mbox{[Fe/H]}\sim-2.5$ in our sample
suggest that at least a few supernovae were responsible for the early
enrichment of these galaxies. Perhaps the supernovae that produced
most of the Fe in UMa\,II and ComBer had (slightly) different
progenitor masses, so that several generations of short-lived stars
contributed different (groups of) elements to the ISM of these
systems. If only a fraction (or even just one) of those SNe hosted
nucleosynthesis through the weak-r process, that could explain the
extremely low and roughly constant levels of Sr and Ba as a function
of [Fe/H] in these systems.

The two Her stars observed by \citet{koch_her} have non-detectable Ba
lines with upper limits of
$\mbox{[Ba/Fe]}<-2.1$. \citeauthor{koch_her} speculated that massive
($\sim35$\,M$_{\odot}$) stars were responsible for the observed light
element abundance pattern.  However, this hypothesis requires that
those massive stars did not produce significant quantities of Ba (or
perhaps any Ba at all), again suggestive of a weak-r signature. Given
such low Ba abundances, the weak r-process would produce a low Ba/Sr
ratio of $\mbox{[Ba/Sr]} \sim -1$ (Figure \ref{srba}).  Thus, although
\citet{koch_her} did not observe Sr, we predict that the Sr abundances
in the Her stars should be relatively large and enhanced with respect
to Ba, and hence potentially measurable. Observations of
neutron-capture elements for additional stars in this and other dwarf
galaxies will shed light on the weak-r hypothesis and will help
disentangle the somewhat peculiar chemical nature of Her, as well as
that of the broader ultra-faint dwarf galaxy population.

\subsubsection{UMa\,II-S3 - an s-rich binary star system?}\label{rs}
Since we observed UMa\,II-S3 with an additional, bluer spectrograph
setting, we were able to obtain abundances of several additional
elements that have strong absorption lines blue-ward of
4150\,{\AA}. Due to the lower $S/N$ ratio in this region ($\sim10$ at
4000\,{\AA}), all these abundances have slightly larger
uncertainties. We derive an Al abundance of
$\mbox{[Al/Fe]}=-0.34\pm0.3$ from the two lines at 3944\,{\AA} and
3961\,{\AA}. This value agrees very well with Cayrel et al. (2004)
halo stars. The Si line at 4102\,{\AA} is very strong and yielded
$\mbox{[Si/Fe]}=0.91\pm0.3$. This is slightly above the trend of the
Cayrel et al. stars, although our value is somewhat uncertain because
the low $S/N$ ratio data hampered the continuum placement. Four Co lines
could also be detected in the bluer setting. Our value of
$\mbox{[Co/Fe]}=-0.09\pm0.1$ agrees well with the other halo star
abundances. We also co-added this bluer spectrum with the spectrum
taken with the ``standard'' setting. This yielded the detection of Zr
at $\lambda$4209. The $\lambda$4317 line was not detected but the
upper limit is consistent with the Zr abundance of
$\mbox{[Zr/Fe]}=-0.60\pm0.3$. The Sr line at 4077\,{\AA} is very strong
and its abundance is consistent with that of the 4215\,{\AA}
line. Because of the higher $S/N$ ratio, we adopt the Sr abundance of
the $\lambda$4215 line. Eu is still not detected in the combined
spectrum. The lines at 4129\,{\AA} and 4205\,{\AA} have similar upper
limits (note that the $S/N$ ratio at 4200\,{\AA} is slightly larger
[$S/N\sim17$] than at 4100\,{\AA} [$\sim13$]). The limit of
$\mbox{[Eu/Fe]} < -0.57$ is also consistent with the $\lambda$4435
line, which is mostly a blend of Ca and Eu.

\begin{figure}
 \begin{center}
   \includegraphics[clip=true,width=8.5cm,bbllx=37, bblly=404,
   bburx=525, bbury=738]{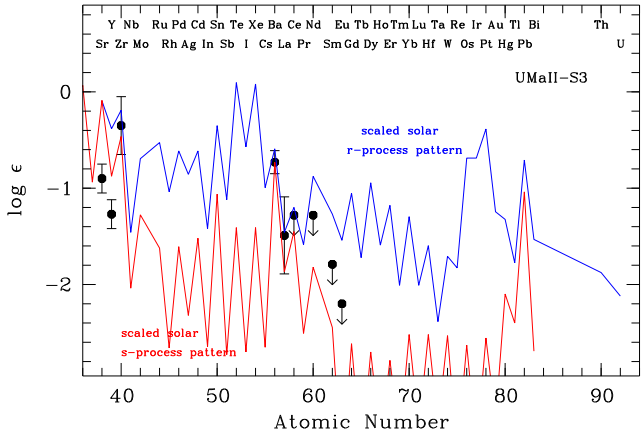} \figcaption{ \label{sproc}
   Neutron-capture abundances and upper limits for UMa\,II-S3
   overplotted with the scaled solar r- and s-process patterns (scaled
   to Ba). There is no agreement with the r-process pattern.  The
   chemical signature of UMa\,II-S3 may, however, be consistent
   with the s-process.  }
 \end{center}
\end{figure}

In summary, we detect not only Sr and Ba in UMa\,II-S3 but also the
neutron-capture elements Zr, Y and La. The latter two were detected in
the redder spectrum. In Figure~\ref{sproc}, we compare our abundances
and upper limits of the neutron-capture elements to those predicted
from the scaled solar r- and s-process patterns \citep{2000burris} to
obtain clues as to where the neutron-capture elements originate. As
can be seen in the figure, our abundances do not agree with the
r-process pattern (particularly the upper limits for Ce, Nd, Sm, and
Eu), but they might be consistent with the s-process pattern.

In the MW halo, s-process-rich metal-poor stars are found down to
metallicities of $\mbox{[Fe/H]}\sim-2.6$ \citep{simmerer2004}.  These
stars experienced mass transfer from a slightly more massive
($\sim1-5$\,M$_{\odot}$) companion as it passed through the AGB phase.
During the mass transfer event, these s-elements, together with
dredged-up C, are donated to the lower-mass companion that we observe
today as a metal-poor giant. Thus, the s-process enrichment is usually
accompanied by an overabundance in C. However, in the case of
UMa\,II-S3, the C-enrichment (often $\mbox{[C/Fe]}>+1.0$) is
missing. Also, the overall neutron-capture abundances in the star are
extremely low, in contrast to that of the usual s-material
overabundances of $\mbox{[s/Fe]}>0$ observed in known s-enriched
metal-poor stars. This fact is rather puzzling and may potentially
challenge the conjecture that UMa\,II-S3 is an s-rich star. If the
star did not receive the s-rich material from a companion, then maybe
a previous generation of AGB stars producing s-elements could have
enriched the gas cloud, e.g., through extensive mass loss, from which
UMa\,II-S3 later formed. Nevertheless, radial velocity variations are
observed for UMa\,II-S3 that indicate binarity, and offer some
observational support for the mass transfer event. Concerning the
binary nature, we have three radial velocity measurements that
demonstrate that this star is indeed in a binary system. We list two
of those values in Table \ref{Tab:obs}, showing a velocity change of
more than 15\,km\,s$^{-1}$ over the course of a year. Note that the
DEIMOS spectrum was taken on 2007 February 13, with $S/N =116$. Its
lower resolution of $R=6,000$ results in a somewhat larger
uncertainty. The third measurement from 2007 November 5 is
$-118.4\pm0.5$\,km\,s$^{-1}$. It was obtained from an $R=50,000$, low
$S/N$  HIRES early test spectrum. Future radial velocity
measurements are required to establish the orbital paramters of this
object.

An unusual chemical history for this star may also explain why other
light elements deviate from the abundances of our other program stars
as well as those of the halo stars. It is somewhat Mg- and Ca-rich
($\mbox{[Mg/Fe]}\sim+0.8$ and $\mbox{[Ca/Fe]}\sim+0.5$) and thus lies
slightly above the general trends seen in our data set (see
Figure~\ref{mgcati}). The Ba/Fe ratio is much higher in UMa\,II-S3
compared with the other stars in the ultra-faint dwarfs, and almost as
high as the Ba abundance found in the more luminous dSphs
\citep{shetrone03, shetrone01} at higher Fe abundances. Interestingly,
this star seems to bridge the gap between our low Ba stars (and also
Dra\,119 and Her) with the stars in the brighter dSphs. This could
also suggest that there is a smooth metallicity \mbox{(Fe-)} dependent
transition.  At the lowest metallicities, Ba (and also Sr) seems to
trail the lower envelope of the halo star abundances, as is also true
for the two most metal-deficient stars from \citet{shetrone03}. Above
$\mbox{[Fe/H]}\sim-2.3$, the halo abundance scatter considerably
tightens and the stars in the more luminous dSphs suddenly appear all
in agreement with the halo abundances. Clearly, more stars at
$\mbox{[Fe/H]}<-2.0$ are required to assess in greater detail whether
the neutron-capture elements in dwarf galaxies generally lie below the
halo, or if UMa\,II-S3 shows an unusual, atypical chemical pattern.
Finally, we stress that any s-process mass transfer onto UMa\,II-S3
should not have affected its light element (Na to Zn) signature, so
this star can still be employed for tracing the chemical composition
of the ISM at the time of its formation.

\section{Chemical History of the \\Ultra-Faint Dwarf Galaxies}\label{sec:history}

Based on the individual abundances of their member stars, we now
discuss the implications of our results for the early chemical history
and star formation history of UMa\,II and ComBer. We also consider how
the ultra-faint dwarf galaxies may fit into the broader picture of
hierarchical galaxy formation.

\subsection{Existence of Extremely Metal-Poor Stars}

\citet{kirby08} presented the first evidence for the existence of
extremely metal-poor stars with $[\mbox{Fe/H}] < -3$ in any dwarf
galaxy, identifying 15 such stars in the ultra-faint dwarf galaxies.
After obtaining high-resolution spectra of two of these targets with
the lowest metallicities (UMa\,II-S1 and UMa\,II-S2), and a
comprehensive uncertainty analysis, the measured Fe abundances are
$\mbox{[Fe/H]}=-3.1$ and $\mbox{[Fe/H]}=-3.2$, respectively (where
$\mbox{[Fe/H]}=\mbox{[Fe\,I/H]}$ = $\mbox{[Fe\,II/H]}$). The total
uncertainties as listed in Table~\ref{err} are 0.3\,dex for Fe\,I and
0.2\,dex for Fe\,II for such giants. From the model atmosphere
comparison in \S~\ref{model_atm_err}, we furthermore find that the
inclusion of scattering would \textit{lower} the measured Fe abundances by
0.1\,dex. In summary, these assessments strongly suggest that within
the uncertainties, the two stars cannot be significantly above
$\mbox{[Fe/H]}=-3.0$, and are in fact true extremely metal-poor stars.
Our results  thus confirm the conclusions of
\citeauthor{kirby08} and suggest that their method is indeed
suitable for identifying EMP stars.

While the sample of stars we have investigated is small, our HIRES
targets were selected only on the basis of their apparent magnitude;
therefore, finding two stars with $[\mbox{Fe/H}] < -3$ out of just six
targets hints that a significant fraction of the stars in the faintest
dwarf galaxies may have had extremely low metallicities.  Previous studies
using both the Ca triplet and high resolution spectroscopy in the
classical dSphs had only identified a handful of stars below $[\mbox{Fe/H}]
= -2.7$ (e.g., \citealt{fulbright_rich,sadakane04,cohen09,aoki09}),
with the vast majority at metallicities above $[\mbox{Fe/H}] = -2$
(e.g., \citealt{shetrone01, shetrone03, koch08a}).  However, since the
total number of stars observed at high resolution across all of the
brighter dSphs is just $\sim50$, it is not yet clear whether the absence of
extremely metal-poor stars in those galaxies reflects a true deficit
or a bias in the Ca triplet [Fe/H] values at low metallicities.

\subsection{Large Internal Abundance Spreads}

Earlier medium-resolution spectra showed that each of the ultra-faint
dwarf galaxies contain stars with a range of Fe abundances
covering $\sim0.5$\,dex rather than a single stellar population
\citep{martin07a, SG07, kirby08, norris_boo}.  Our measurements conclusively
demonstrate that these abundance spreads are real --- even with only
three stars in each galaxy, our targets span a range of 0.9\,dex in
UMa\,II and 0.6\,dex in ComBer.

There are several ways to produce the internal abundance spreads
observed in these extremely low luminosity galaxies.  (1) If the stars
were formed in multiple smaller progenitor systems that later merged
to become the dwarf galaxy we see today, then it would be natural for
the ISM in each of the proto-dwarf galaxies to have had a different
metallicity.  (2) If the stars formed in situ (i.e., in the main halo
that became the present-day dwarf galaxy), the star-forming gas may
have been incompletely mixed either as a result of asymmetric
supernova explosions or rapid star formation before mixing could
occur.  (3) Finally, if the young UMa\,II and ComBer were able to hold
onto their gas for an extended period of time (or re-accrete enough
gas later to produce multiple epochs of star formation), the stars
formed at later times would have higher metallicities because of the
continued chemical evolution of the ISM.

One way to separate these possibilities may be high signal-to-noise
photometry in the main sequence turnoff region to determine whether
there is a significant age spread among the member stars.  Obtaining
spectra of more metal-rich stars in these two dwarf galaxies would
also be useful to see if the $[\mbox{$\alpha$/Fe}]$ ratios indicate
any contribution from Type Ia supernovae.  Our data are consistent
with a constant $\mbox{[$\alpha$/Fe}]$ as a function of [Fe/H], but
measurements at $\mbox{[Fe/H]} \ge -2$ are needed to reveal whether
the turn-down in $\mbox{[$\alpha$/Fe]}$ seen in systems with extended
star formation histories (e.g., the classical dSphs and the MW halo)
is present in the ultra-faint dwarf galaxies as well.

\subsection{Comparison to the Milky Way Halo \\and the More Luminous dSphs}
\label{sec:halocomparison}

Contrary to previous studies of the more luminous dSphs (e.g.,
\citealt{shetrone01, shetrone03, tolstoy03, sadakane04}\footnote{Their
  lowest metallicity star shows a somewhat similar chemical pattern to
  that found in the present study indicating SN\,II enrichment. The
  large Fe spread in UMi of $\sim1$\,dex is comparable to what is
  found in UMa\,II, although reaching higher Fe values and
  reflecting enrichment by SNe Ia as well as massive stars.};
\citealt{venn04, geisler07}, and see \citealt{tolstoy_araa} for a
recent review on this topic), the abundances found in our two
ultra-faint dwarf galaxies generally agree with those of MW halo stars
(see Figure~\ref{cayrel_abundances_light}).  The observed abundance
pattern includes the halo-typical abundance offset of $\sim0.4$\,dex
among the $\alpha$-elements that results from ISM enrichment by
massive Type\,II supernovae (e.g.,
\citealt{woosley_weaver_1995}). This agreement may point to an initial
mass function similar to the one that produced the halo star abundance
pattern at early times, and furthermore to a significant contribution
from massive stars to the early enrichment of the ultra-faint
systems. This would be consistent with the overall low metallicity of
all of the newly discovered dwarf galaxies \citep{kirby08}, and
feedback from those supernovae might be responsible for suppressing
star formation in these systems. The more luminous dSphs have
significantly higher average metallicities than the ultra-faint dwarf
galaxies. Their lower, more solar-like $\alpha$-element ratios (see
Figure~\ref{mgcati}) clearly point to a major contribution of Fe from
Type Ia supernovae, and potentially a mass function shifted to lower
masses. \citet{tolstoy03} pointed out that small systems may naturally
have an initial mass function with a suppressed high mass end as a
result of the difficulty of forming large molecular clouds (and
therefore very massive stars) in such low density environments.

Our two most metal-poor stars (at $\mbox{[Fe/H]}<-3.0$) are enriched
in carbon. In the halo, there is an increasing trend of C enhancement
with decreasing metallicity. Although our sample is quite
small, this may be another signature that the ultra-faint dwarf galaxies share
with the MW halo. Overall, C excesses in the most metal-poor stars
point toward the important role of C in the early universe and as
a potential cooling agent of the primordial ISM
\citep{ARAA,aoki_cemp_2006,dtrans}. The presence of C-rich extremely metal-poor
stars is consistent with the assumption that the massive stars in a
system produced the C either during stellar evolution or during their
supernova explosions.

It is less clear whether the stellar abundances of neutron-capture
elements in the ultra-faint dwarf galaxies are consistent with those
in the MW halo and more luminous dSphs. The observed Ba and Sr
abundances are near the low end of the neutron-capture-to-iron ratios
seen in halo stars, whereas the abundances found in the more luminous
dSphs (at higher metallicity) agree rather well with those found in
the halo. Low neutron-capture abundances are also found in Hercules
\citep{koch_her}, two stars in the more luminous dSph Draco
\citep{fulbright_rich, shetrone01}, one star in Sextans
\citep{shetrone01}, and one in \citet{sadakane04}. An important
question therefore arises as to how a system can enrich itself
significantly with iron-peak elements, but produce very little to no
neutron-capture elements in a consistent way over a long period of
time (up to relatively high metallicities of
$\mbox{[Fe/H]}\sim-2.3$). Our observations underscore that the
production of these groups has to be strongly decoupled, as already
evidenced by halo star abundance patterns (e.g.,
\citealt{sneden_araa}). This question can only be adequately addressed
with a much larger sample of stars in the ultra-faint dwarf galaxies,
and detections of heavier neutron-capture species would also be
helpful. In the meantime, we speculate in Section~\ref{heavy} that the
neutron-capture elements in the ultra-faint dwarfs were produced
through the weak r-process in stars more massive than
$\sim8-10$\,M$_{\odot}$ (which is the mass range of the main
r-process, see \S~\ref{nc}).  Such massive stars
($\sim20$\,M$_{\odot}$; e.g., \citealt{wanajo_ishimaru}) might also
have been responsible for the halo-like levels of $\alpha$-elements
observed in our stars. \citet{izutani} recently calculated weak
r-nucleosynthesis yields and found that energetic hypernovae with
20\,M$_{\odot}$ are needed to reproduce the supposed weak r-abundance
pattern in metal-poor halo stars. This agrees with findings by
\citet{nomoto2006}, who explained the abundances of the
\citet{cayrel2004} stars with their hypernova models. Since our stars
are very similar to the Cayrel et al.~stars (see
Figure~\ref{cayrel_abundances_light}) this is a plausible explanation
for the chemical signature of stars with $\mbox{[Fe/H]}\lesssim-2.5$.
Because our abundance measurements point to the weak r-process,
operating in massive stars, as the source of the heavy elements in the
ultra-faint dwarfs, it might even be the case that the population of
lower mass ($\sim8-10$\,M$_{\odot}$) SN\,II progenitors was suppressed
at early times in these objects.  In summary, the low levels of
neutron-capture elements may simply reflect a local environment that
was driven by a particular mass function of its stars.

These conclusions regarding the source of the r-process enrichment
differ strongly from the conclusions of previous
studies. \citet{venn04} suggested that in the more luminous dSphs
there may be a lack of hypernovae and \citet{tolstoy03} suggested that
the IMF was shifted to lower mass stars. However, given the different
abundance patterns of the ultra-faint dwarfs and the more luminous
dSphs this contradiction may not be surprising. Leaving aside all the
uncertainties in nucleosynthesis processes and small-number statistics
of our small samples, the abundance data themselves (see
Figures~\ref{mgcati} and \ref{srbafe}) reveal different enrichment
histories for the ultra-faint dwarfs and their brighter counterparts.
This contrast between the abundance patterns seen in high and low
luminosity dwarf galaxies demonstrates again that the ultra-faint
dwarfs cannot simply be tidally stripped versions of the classical
dSphs \citep{penarrubia08,kirby08,geha08a}.  Spectroscopy of more stars
in the ultra-faint dwarf galaxies, as well as in the more luminous
dSphs, is needed to establish more firmly exactly how the Milky Way's
population of dwarf galaxies evolved and to what extent their chemical
abundances are correlated strictly with luminosity. As \citet{kirby08}
showed, there is a strong correlation of Fe abundance with luminosity
but for other elements this picture may be different.

While we were in the process of completing this paper, two new studies
of brighter dSphs were published. \citet{cohen09} presented a
high-resolution abundance analysis of 8 stars in the classical dSph
Draco. Their stars span the range from $-3.0\leq\mbox{[Fe/H]}\leq-1.5$,
with one star at $\mbox{[Fe/H]}=-3.0$. At the low-metallicity end they
find the abundances of several elements to be in agreement with those
of halo stars, but at higher metallicities, deviations are found.  The
$\alpha$-abundances (Ca and Ti), however, are depleted relative to the Milky Way halo,
as has been found in the higher metallicity stars in the luminous
dwarf galaxies. The Sr abundances at the low-metallicity end are
similarly low as found in the ultra-faints, but rise up to solar at
higher metallicities. The Ba values follow a similar trend but at a
slightly more elevated level.

In Sextans, \citet{aoki09} found one star with $\mbox{[Fe/H]}=-3.1$
and and five with $-2.9<\mbox{[Fe/H]}<-2.7$. Their most metal-poor
star seems to mostly follow the Galactic halo [X/Fe] trends in the
same fashion as the objects presented in this study.  However,
the Ba/Fe ratio in their $\mbox{[Fe/H]}=-3.1$ star is high
($\mbox{[Ba/Fe]}=0.5$), which is different from what has been found in
this study. On the other hand, their slightly more metal-rich stars
show depletions similar to those generally found in the luminous dwarf
galaxies (at $\mbox{[Fe/H]}\gtrsim-2.5$), as well as the low Ba
abundances that have been found for our UMa\,II and ComBer stars.

These new results suggest that there may be a metallicity ([Fe/H])
dependence for elemental ratios to be more halo-like at metallicities
below $\mbox{[Fe/H]}=-2.5$ in all dwarf galaxies, not just the
ultra-faint ones. However, the body of stellar data in the luminous
dwarf galaxies may not yet be sufficient to derive strong conclusions
about their most metal-poor stars, especially in light of the fact
that counter examples are also evident (e.g., low Ca and Ti at
$\mbox{[Fe/H]}=-3.0$; \citealt{cohen09}).

\subsection{Comparison to Globular Cluster Abundances}\label{gc}

We have so far found evidence that the chemical signatures of stars in
the ultra-faint dwarf galaxies closely resemble those of halo field
stars. An outstanding question, then, is if there are also
similarities to globular cluster stars. It has been debated, and so
far been excluded (e.g., \citealt{SG07}), that the ultra-faint dwarf
galaxies might be globular clusters instead of dark matter-dominated
galaxies. Adding chemical information to this discussion may provide
further constraints on the origin of these dim systems. A first and
obvious difference between the dwarf galaxy stars and the globular
cluster members is the large Fe abundance spread compared with the
mono-metallic populations in globular clusters. Furthermore, our Fe
values are generally lower than those of the most metal-poor globular
cluster \citep{harris97a}. This behavior was already pointed out by
\cite{SG07} and \citet{kirby08} for the ultra-faint dwarf galaxies and
is not limited to the two systems studied here.

Aside from the very different behavior in Fe abundances, there exist a
number of additional characteristic globular cluster abundance
patterns involving low C and O, and high N, Na, and Al that we can
investigate in the ultra-faint dwarf galaxies.

For the low [C/Fe] globular cluster giants, it has been shown (e.g.,
\citealt{shetrone_gc_ch}) that deep mixing events on the red giant
branch and prior nucleosynthesis could both be responsible for the low
C values, but these mechanisms are difficult to distinguish even in
extreme cases (e.g., \citealt{sneden2004}).  It is thus unclear if the
level of depletion in cool giants arises from deep mixing where
material in which C has been converted to N is dredged up to the
surface or if the stars were simply born from C-poor material since
the C-N anticorrelation is found down to main-sequence stars in some
clusters \citep{harbeck_gc}.  Since we have no N and O measurements,
so we are unfortunately not able to address in detail whether the low
C abundances observed in our dwarf galaxy stars arise from
carbon-depleted material. Irrespective of considering our stars as
globular cluster or halo star analogs, their evolutionary status alone
suggests that some mixing may already have taken place and the
currently observed subsolar levels of two of our stars are in
agreement with halo stars of similar evolutionary status (and which
are assumed not to have formed from C-depleted material). This
suggests that our stars did not form from particularly C-poor gas as
some clusters have. The only firm conclusion that is possible from the
available C abundances alone is that our two stars with the highest C
values ($\mbox{[C/Fe]}\sim +0.5$ and $\sim+0.8$; both in UMa\,II)
appear to be more carbon-rich than is typical in clusters.  And even
the two stars with $\mbox{[C/Fe]}\sim 0.0$ (both in ComBer) do not
necessarily indicate either substantial mixing or being born from very
C-depleted material. We thus conclude that at present there is no
strong indication that the C abundances in dwarf galaxies behave
similarly to those in globular clusters (where the C-depletion is not
due to deep mixing alone).

For the following discussion, we correct our Na abundances for non-LTE
effects for the comparison with the results of \citet{sneden2004}. We
thereby adopt the same corrections \citep{gratton99_nlte} as Sneden et
al.  For completeness, we remind the reader that in the earlier
comparison with the Cayrel et al. halo sample, the uncorrected LTE Na
values were used.
\\ \textit{Na-O correlation:} While we have no O abundances available,
we can nevertheless consider the Na abundance distribution of globular
cluster stars and halo field stars. In clusters, it has been found
that the stars closest to the red giant branch tip have the highest Na
abundances (up to $\mbox{[Na/Fe]}\sim1.0$; \citealt{sneden2004} and
references therein). This may originate from proton-capture and thus
be a sign of deep mixing. The halo appears to lack such extreme
equivalents. If the mixing scenario is correct, we should not find
such extreme Na abundances in our sample since our targets do not sit
at the tip of the giant branch. Indeed, our corrected Na/Fe values,
span the range of $-0.45$ to $+0.45$\,dex. There is an overlap in Na
abundances between the ultra-faint dwarf galaxies and globular
clusters, although the former systems also have some stars with Na
abundance lower than those ever seen in clusters.
Regarding the one star with
$\mbox{[Na/Fe]}\sim0.45$ (ComBer-S3), it is unclear whether this could
be interpreted as some globular cluster signature because the star
with the lowest Na, $\mbox{[Na/Fe]}\sim-0.7$, is in the same dwarf
galaxy (ComBer\,II-S1). Such low abundances are not found in
clusters.   \\ \textit{Na-Al
  correlation:} We have one Al detection available, yielding a low
abundance ($\mbox{[Al/Fe]}=-0.34$). Together with the corresponding Na
abundance, UMa\,II-S3 does not lay on the Na-Al correlation for
clusters presented in \citet{sneden2004}. The lack of additional Al
measurements precludes a strong conclusion, although the one detection
further supports the idea that UMa\,II shows no distinct sign for a
cluster chemical history.\\ \textit{Na-Mg correlation:} All six of our
targets (as well as HD122563) lie in the high-Mg, low-to-intermediate
Na range of Figures 13 and 14 of \citet{sneden2004}. These authors
show that this region is sparsely and in some cases not at all
populated by cluster members. We interpret this as an additional clue
to the non-cluster-like chemical origin of our targets.

In summary, the significant difference of Fe abundances and spreads
between clusters and ultra-faint dwarfs already suggested that our
stars are not associated with globular clusters or globular
cluster-like nucleosynthesis histories and events.  Based on our
examination of the abundance correlations typically seen in globular
clusters, we do not find clear signs for any correspondence between
the abundance patterns in the ultra-faint dwarfs and those of globular
clusters. We note, though, and given our limited sample size, more
observations of ultra-faint dwarf galaxy stars could be helpful to
further address this issue.

\subsection{Building Up the Metal-Poor Halo}

The similarity of the abundance pattern we find in UMa\,II and ComBer
(\S \ref{sec:lightelements} and \ref{sec:halocomparison}) to the
well-known abundance pattern seen in very low-metallicity MW halo
stars suggests the possibility of a common origin of the two
populations.  Could the bulk of the metal-poor end of the halo
metallicity distribution have been formed in galaxies similar to the
ultra-faint dwarfs?

\citet{kirby08} demonstrated that integrated over the ultra-faint
dwarf galaxies with $-8 \lesssim M_{V} \lesssim -4$, $\sim5$\,\% of
the stars have metallicities $\mbox{[Fe/H]} \le -3$.  Our results
suggest that for the lowest luminosity galaxies the fraction of EMP
stars may be even higher, but for galaxies with stellar masses of
$\sim10^{4}$\,M$_{\odot}$ \citep{martin08} $\sim5$\,\% should be a
representative value.  Every such galaxy that has been destroyed by
the Milky Way therefore must have added $\sim500$\,M$_{\odot}$ of EMP
stars to the halo.

The fraction of such low metallicity stars in the brighter dSphs is
not currently known, since no stars with $\mbox{[Fe/H]} \le -3$ have
been detected in those galaxies, but the large number of stars
observed without finding any EMP stars \citep[e.g.,][]{helmi06,koch06}
indicates that $\sim0.1$\,\% would be a reasonable assumption.  Given
a typical stellar mass of $\sim10^{6}$\,M$_{\odot}$ (similar to
Sculptor, Carina, or Sextans), the destruction of a classical dSph
would add $\sim1000$\,M$_{\odot}$ of EMP stars to the halo.

Current estimates are that the Milky Way's satellite population today
includes $\sim5$ times as many ultra-faint dwarf galaxies as classical
dSphs \citep{SG07, koposov08, tollerud08}.  If the population of
dwarfs that has been cannibalized over the lifetime of the Milky Way
had a similar luminosity function, then the progenitors of the
ultra-faint dwarfs would have provided $\sim2.5$ times as many EMP
stars as the more luminous dSphs.  Thus, even though the ultra-faint
dwarf galaxies simply do not contain enough stars to contribute
significantly to the overall stellar mass of the MW halo, it is
plausible that such galaxies could be the source of a large fraction
of the most metal-poor halo stars.

\subsection{$\Lambda$CDM Simulations}
We now consider whether the observational results summarized in \S4.1
--- 4.3 are consistent with expectations from the predictions of
$\Lambda$CDM plus galaxy formation models, where the Milky Way's halo
is largely composed of stars stripped from infalling dwarf galaxies.

The most detailed set of predictions have come from the semi-analytic
plus N-body model described in \citet{bullock05a} and
\citet{robertson05} (see also the related papers by \citealt{font06a}
and \citealt{johnston08a}.)  The discrepant $\alpha$-abundance
patterns of stars in the classical dSphs and the Milky Way's field
halo population are consistent with this model, owing to two important
differences between the surviving MW dwarf galaxies and those that
were destroyed to build the stellar halo. First, the stars in the
halo came from dwarf galaxies accreted on average 9 Gyr in the past,
whereas the dwarf galaxies surviving today were accreted on average 5
Gyr ago.  Second, the majority of stars in the halo were formed in
dwarf galaxies substantially more massive ($M \sim 5 \times 10^{10}$
$M_{\odot}$) than the dwarf satellites surviving around the Milky Way
until today ($M \sim 5 \times 10^8$ $M_{\odot}$).

This model is broadly consistent with our observations that the
ultra-faints have (1) a high frequency of extremely metal-poor stars
(see also \citealt{kirby08}), and (2) abundance patterns generally
consistent with those of Milky Way field halo stars.  The chemical
evolution model of \citet{robertson05}, \citet{font06a}, and
\citet{johnston08a} predicts that, although massive dwarf galaxies are
the source of the vast majority of the mass in the stellar halo, low
luminosity, metal-poor, galaxies accreted at early times contribute
some stars to the metal-poor stellar halo and have [$\alpha$/Fe]
abundances similar to those observed in low metallicity halo stars.
\citet{robertson05} emphasize that the lowest [Fe/H] stars even in
massive dwarf galaxies (such as the Magellanic Clouds, and the dSphs
predicted to build up the majority of the halo) should also exhibit
similar abundance patterns to the most metal-poor halo stars, although
this prediction has not yet been tested by observations.

In an alternative approach, \citet{prantzos08} uses an analytic model
to predict the metallicity distribution function (MDF) of the Milky
Way halo from a set of satellites with stellar masses ranging from $2
\times 10^6$ to $2 \times 10^8$ M$_{\odot}$.  In his model, the shape
of the MDF is also a strong function of satellite mass, with lower
stellar mass satellites having a much higher fraction of low
metallicity stars than the higher mass galaxies, consistent with the
findings of \citet{kirby08}.  He concludes that low mass systems are
expected to contribute substantively to the low-metallicity tail of
the global Milky Way halo MDF.

While our results seem to fit naturally into this picture, we note
that the lowest mass Milky Way satellites in the models of both groups
have over an order of magnitude more stellar mass than the ultra-faint
dwarfs, and models including lower luminosity systems are therefore
still needed.  However, the predicted trends between dwarf galaxy
mass/luminosity, accretion time, [Fe/H] and [$\alpha$/Fe]
\citep{bullock05a,robertson05,johnston08a,prantzos08} can still be used
to infer that our observational results are broadly consistent with
the presently favored cosmological model.

\section{Concluding Remarks}\label{sec:conc}

We have presented a high-resolution chemical abundance analysis for
six stars located in the ultra-faint dwarf galaxies Ursa Major~II and
Coma Berenices.  Comparing these results with previous studies of
stars in the Milky Way halo and the brighter, classical dSphs, we
arrive at the following conclusions:

\begin{itemize}
\item{\textit{The ultra-faint dwarf galaxies contain significant
      numbers of extremely metal-poor stars with $\mbox{[Fe/H]} <
      -3$.}  Two out of the six stars in our sample (both in UMa\,II)
    fall into this category, despite our metallicity-independent
    selection criteria.  Although the sample is obviously small, the
    statistics are inconsistent with the metallicity distributions of
    both the brighter dSphs and the MW halo.  \citet{schoerck} find
    that less than 2\% of halo stars are at $\mbox{[Fe/H]} < -3$, so
    the probability of our sample containing two such stars is very
    low.  In the brighter dSphs, not a single star out of 47 observed
    at high spectral resolution and several thousand observed at lower
    resolution has been found with $\mbox{[Fe/H]} < -3$.  Thus, not
    only do the ultra-faint dwarf galaxies have lower mean
    metallicities than any other known stellar systems, but they also
    appear to contain a larger fraction of EMP stars. }

\item{\textit{The ultra-faint dwarf galaxies have large internal metallicity spreads.}
  Confirming the earlier results at lower spectral resolution of
  \citet{SG07}, \citet{kirby08}, and \citet{norris_boo}, we find that
  our three stars in each galaxy span a range of $\sim0.9$~dex in
  [Fe/H] in UMa\,II and $\sim0.6$~dex in ComBer.  Again, despite the
  small samples, it is clear that there are significant internal
  metallicity variations in these objects.  The metallicity spreads
  have important implications for the formation of the ultra-faint
  dwarf galaxies, suggesting either early star formation in multiple
  proto-dwarf galaxies that later merged, extended star formation histories, or
  incomplete mixing in the early ISM.  Distinguishing between these
  scenarios requires larger samples of high-resolution spectroscopy of
  stars covering a wider range of metallicities and improved age
  constraints from photometric studies. }

\item{\textit{The abundance pattern of light elements ($Z < 30$) in the
  ultra-faint dwarf galaxies is remarkably similar to the Milky Way
  halo.}  In contrast to what is seen in the brighter dSphs
  \citep[][and references therein]{venn04}, we find that the trends of
  $\alpha$ and iron-peak abundances with [Fe/H] in the ultra-faint
  dwarf galaxies are in excellent agreement with the best halo samples
  over the same metallicity range of $-3.2 \lesssim \mbox{[Fe/H]}
  \lesssim -2.34$.  This result suggests that the metal-poor end of
  the MW halo population could have been built up from the destruction
  of large numbers of systems similar to the ultra-faint dwarf
  galaxies. }

\item{\textit{Neutron-capture elements, specifically Ba and Sr, have extremely
    low abundances in the ultra-faint dwarf galaxies.}  Particularly in
    ComBer, the Ba and Sr values observed are well below the
    abundances found in MW halo stars with similar Fe abundances.  In
    UMa\,II, the neutron-capture abundances are lower than the halo
    averages, but not outside the distribution.  Both galaxies exhibit
    a large scatter ($\sim1$~dex) in the abundances of these elements.
    This abundance pattern may originate from the weak r-process,
    whose site is unknown but is speculated to be in very massive
    stars ($M \gtrsim 20$~M$_{\odot}$). }

\item{\textit{The results above are broadly consistent with the
      predictions of the currently favored cosmological models
  \citep[e.g.,][]{bullock05a,robertson05,johnston08a,prantzos08}.}
  While the majority of the mass in the stellar halo was formed in
  much larger systems, our results support a scenario where galaxies
  similar to the faintest dwarf galaxies may have been the source for
  much of the metal-poor end of the Milky Way halo Fe metallicity
  distribution.}

\end{itemize}

This study has provided the first evidence that the chemical evolution
in two of the faintest dwarf galaxies known may have been similar to
that of the MW halo.  We find some intriguing and yet-to-be-explained
abundance signatures, such as the low neutron-capture abundances, but
overall there is a surprising level of agreement between the
abundances in the ultra-faint dwarf galaxies and the most metal-poor
halo stars (see Figure~\ref{cayrel_abundances_light}). Fully
unraveling the complex relationship between the entire population of
observed dwarf galaxies and the formation of the stellar halo of the
Milky Way, however, will require more spectroscopic and photometric
data.  It is not yet clear whether the differences we have found from
previous studies of the brighter MW dSphs --- the presence of
extremely metal-poor stars in the ultra-faint dwarf galaxies and the
agreement between the light-element abundance pattern in the
ultra-faint dwarf galaxies and the halo --- stem from actual
differences between the classical dSphs and their much fainter cousins
or observational biases.  Spectroscopy of more metal-poor stars in the
bright dSphs and more metal-rich stars in the ultra-faint dwarf
galaxies \citep[e.g.,][Ivans et al. 2008, in prep.]{koch_her} to
increase the sample sizes and broaden the range of overlap between the
two types of galaxies will help  clarify this picture
and provide further clues to the formation of the halo of our
Galaxy.

\acknowledgements{Data presented herein were obtained at the
  W. M. Keck Observatory, which is operated as a scientific
  partnership among the California Institute of Technology, the
  University of California, and the National Aeronautics and Space
  Administration.  The Observatory was made possible by the generous
  financial support of the W. M. Keck Foundation.  The authors wish to
  recognize and acknowledge the very significant cultural role and
  reverence that the summit of Mauna Kea has always had within the
  indigenous Hawaiian community.  We are most fortunate to have the
  opportunity to conduct observations from this mountain. We thank
  Evan Kirby for providing information on the stellar parameters,
  Kjell Eriksson for computing MARCS model atmospheres, Ian Roederer,
  Volker Bromm, and Grant Matthews for enlightening discussions about
  neutron-capture nucleosynthesis, and Chris Sneden, John Norris and
  the anonymous referee for helpful suggestions regarding the
  manuscript. We are especially indebted to Xavier Prochaska for his
  extensive help with running the HIRES data reduction pipeline.
  A.~F. acknowledges support through the W.~J.~McDonald Fellowship of
  the McDonald Observatory and the Clay Fellowship administered by the
  Smithsonian Astrophysical Observatory.  J.D.S. gratefully
  acknowledges the support of a Millikan Fellowship provided by
  Caltech and a Vera Rubin Fellowship from the Carnegie Institution of
  Washington.}

\textit{Facilities:} \facility{Keck I (HIRES)}

\clearpage

\begin{deluxetable}{lllllcr} 
\tablecolumns{10} 
\tablewidth{0pt} 
\tabletypesize{\small}
\tablecaption{\label{Tab:pho} Photometry}
\tablehead{
\colhead{Galaxy} & 
\colhead{SDSS designation} & 
\colhead{Star} & 
\colhead{$V$} & 
\colhead{$r$ } & 
\colhead{$g-r$ } & 
\colhead{$M_{r}$ } }
\startdata
Ursa Major II &SDSS J084954+630822 & UMa\,II-S1 & 18.13 & 17.85 &   0.68 &   0.08\\
Ursa Major II &SDSS J085002+631333 & UMa\,II-NM & 16.83 & 16.54 &   0.70 &$-$1.15\\
Ursa Major II &SDSS J085234+630501 & UMa\,II-S2 & 17.66 & 17.37 &   0.72 &$-$0.38\\
Ursa Major II &SDSS J085259+630555 & UMa\,II-S3 & 16.79 & 16.43 &   0.86 &$-$1.33\\
Coma Berenices&SDSS J122643+235702 & ComBer-S1  & 18.12 & 17.89 &   0.58 &$-$0.29\\
Coma Berenices&SDSS J122655+235610 & ComBer-S2  & 17.50 & 17.24 &   0.65 &$-$0.94\\
Coma Berenices&SDSS J122657+235611 & ComBer-S3  & 18.02 & 17.77 &   0.62 &$-$0.41
\enddata
\tablecomments{UMa\,II-NM was observed but was later determined not to
  be a member of UMa\,II.}
\end{deluxetable} 

\begin{deluxetable}{lrrrccccrrc} 
\tablewidth{0pt} 
\tabletypesize{\small}
\tablecaption{\label{Tab:obs} Observing Details}
\tablehead{
\colhead{Star} & \colhead{$\alpha$}&\colhead{$\delta$}&\colhead{JD}&\colhead{$t_{\rm {exp}}$ } & 
\colhead{S/N}    & \colhead{S/N}    & \colhead{S/N}    & 
\colhead{$v_{\rm{rad,HIRES}}$ }  & \colhead{$v_{\rm{rad,DEIMOS}}$ }  & \colhead{Comment} \\
\colhead{}&\colhead{(J2000)}&\colhead{(J2000)}&\colhead{ }&\colhead{hr}&\colhead{5000\,{\AA}}&\colhead{6000\,{\AA}}&\colhead{6500\,{\AA}}&\colhead{(km\,s$^{-1}$)}&\colhead{(km\,s$^{-1}$)}&\colhead{} }
\startdata
UMa\,II-S1  & 08~49~53.46   & 63~08~21.94  & 2454520.8   & 5.00 & 20  & 30  & 38  &$-124.5\pm 0.3$ &$-121.5 \pm 2.2$ &target  \\
UMa\,II-NM  & 08~50~01.84   & 63~13~33.05  & 2454521.0   & 1.50 & 33  & 33  & 42  &$-112.0\pm 0.2$ &$-111.0 \pm 2.2$ &target  \\
UMa\,II-S2  & 08~52~33.50   & 63~05~01.33  & 2454519.8   & 3.00 & 24  & 37  & 42  &$-110.6\pm 0.3$ &$-107.5 \pm 2.2$ &target  \\
UMa\,II-S3  & 08~52~59.07   & 63~05~54.81  & 2454519.8   & 1.00 & 20  & 27  & 30  &$-119.8\pm 0.2$ &$-102.6 \pm 2.2$ &target  \\
ComBer-S1   & 12~26~43.47   & 23~57~02.47  & 2454522.0   & 5.29 & 22  & 27  & 29  &$  93.8\pm 0.3$ &  $97.3 \pm 2.2$ &target  \\ 
ComBer-S2   & 12~26~55.46   & 23~56~09.83  & 2454520.0   & 5.25 & 23  & 28  & 30  &$  96.4\pm 0.2$ &  $97.5 \pm 2.2$ &target  \\
ComBer-S3   & 12~26~56.67   & 23~56~11.84  & 2454521.1   & 2.83 & 33  & 47  & 51  &$  99.0\pm 0.3$ & $102.6 \pm 2.2$ &target  \\  
HD~122563   & 14~02~31.85   & 09~41~09.94  & 2454521.9   & 0.01 & 530 & 650 & 750 &$ -25.1\pm 0.2$ &      \nodata    &standard
\enddata

\tablecomments{The $S/N$ measurements are for $\sim22$\,m{\AA}
(``blue'' CCD), $\sim26$\,m{\AA} (``green'' CCD) and
$\sim28$\,m{\AA} (``green'' CCD) pixel sizes, respectively.}
\end{deluxetable} 

\begin{deluxetable}{lllrcr} 
\tablecolumns{10} 
\tablewidth{0pt} 
\tabletypesize{\small}
\tablecaption{\label{Tab:merit} Comparison of ``figure of merit'' of literature studies}
\tablehead{
\colhead{Study} & 
\colhead{R} & 
\colhead{$S/N$} & 
\colhead{$\lambda$} & 
\colhead{F} & 
\colhead{Comment}\\
\colhead{} & 
\colhead{} & 
\colhead{[per pixel]} & 
\colhead{[{\AA}]} & 
\colhead{} & 
\colhead{} 
 }
\startdata

\multicolumn{6}{c}{Selected luminous dSphs studies}\\\hline
\citet{shetrone98}    & 34000 & 24--29 & 6300 & 130--157 &    Draco\\
\citet{shetrone01}    & 34000 & 24     & 6100 & 134      &      Draco \\
\citet{shetrone01}    & 34000 & 19--36 & 6100 & 106--210 &    Ursa Minor\\
\citet{shetrone01}    & 34000 & 13--27 & 6100 &  72--150 &    Sextans \\
\citet{shetrone03}    & 40000 & 30     & 5800 & 207      &  Scl, Fnx, Car, Leo \\
\citet{bonifacio04}   & 43000 & 19--43 & 5100 & 160--363 &    Sagittarius\\
\citet{sadakane04}    & 45000 & 50--60 & 6100 & 368--442 &    Ursa Minor\\
\citet{geisler05}  & 16000 & 65     & 4500 & 231 &    Sculptor\\
\citet{geisler05}  & 22000 & 120    & 6700 & 394 &    Sculptor\\ 
\citet{koch_her}      & 20000 & 32     & 6500 &  98      &   Hercules \\\hline\\

\multicolumn{6}{c}{This study}\\\hline
Ursa Major\,II    & 34000 & 12--15 & 4500 & 91--136 &    \\
Ursa Major\,II    & 34000 & 20--24 & 5000 & 136--163 &    \\
Ursa Major\,II    & 34000 & 30--42 & 6500 & 157--220 &    \\
Coma Berenices     & 34000 & 11--23 & 4500 & 83--174  &    \\
Coma Berenices     & 34000 & 22--33 & 5000 & 150--224 &    \\
Coma Berenices     & 34000 & 29--51 & 6500 & 152--267 &    
\enddata	 		 
\end{deluxetable}

 \begin{deluxetable}{lllrrrrrrrrrrrrrrr}
\tablecolumns{8}
\tablewidth{0pt}
\tabletypesize{\footnotesize}
\tabletypesize{\tiny}
\tablecaption{\label{Tab:Eqw} Equivalent width measurements}
\tablehead{
\colhead{El.} & \colhead{$\lambda$} &\colhead{$\chi$} &\colhead{$\log gf$}&
\colhead{EW}&\colhead{$\lg\epsilon$}&
\colhead{EW}&\colhead{$\lg\epsilon$}&
\colhead{EW}&\colhead{$\lg\epsilon$}&
\colhead{EW}&\colhead{$\lg\epsilon$}&
\colhead{EW}&\colhead{$\lg\epsilon$}&
\colhead{EW}&\colhead{$\lg\epsilon$}&
\colhead{EW}&\colhead{$\lg\epsilon$}\\
\colhead{}&\colhead{}&\colhead{}&\colhead{}&
\multicolumn{2}{c}{UMa\,II-S1}&
\multicolumn{2}{c}{UMa\,II-S2}&
\multicolumn{2}{c}{UMa\,II-S3}&
\multicolumn{2}{c}{ComBer-S1}    &
\multicolumn{2}{c}{ComBer-S2}    &
\multicolumn{2}{c}{ComBer-S3}    &
\multicolumn{2}{c}{HD122563}  
}
\startdata
CH    &  4313    & \nodata &  \nodata  &   syn   &    6.01 &      syn  &      5.76 &     syn  &    5.21 &   syn  &   6.06 &   syn  &   5.61 &    syn  &  5.31 &    syn  &   5.52\\ 
CH    &  4322    & \nodata &  \nodata  &   syn   &    6.10 &      syn  &      5.93 &     syn  &    5.26 &   syn  &   6.11 &   syn  &   5.58 &    syn  &  5.53 &    syn  &   5.51\\
O\,I  &  6300.31 &    0.00  & $-$9.75  &$<$15.0  & $<$7.30 & $<$14.0   &   $<$6.84 &$<$14.0   & $<$6.96 & $<$12.0& $<$7.07&  22.6  &   7.34 & $<$11.0 &$<$6.85& \nodata & \nodata \\
O\,I  &  6363.79 &    0.02  & $-$10.25 &\nodata  &\nodata  &   \nodata &   \nodata & \nodata  & \nodata & \nodata&\nodata &\nodata &\nodata & \nodata &\nodata&    1.9  &  6.40  \\
Na\,I &  5889.95 &    0.00  &    0.101 &   124.0 &   3.12  &    128.5  &     2.84  &   197.0  &   3.64  &  151.7 &   3.13 &\nodata &\nodata &   194.3 &  3.76 &   187.3 &  3.49 \\ 
Na\,I &  5895.92 &    0.00  & $-$0.197 &   102.8 &   3.06  &    117.0  &     2.96  &   171.3  &   3.64  &  138.2 &   3.23 &  132.9 &   3.21 &   179.1 &  3.89 &   171.8 &  3.59 \\ 
Mg\,I &  4571.10 &    0.00  & $-$5.688 &    32.3 &   4.96  &     40.4  &     4.82  &   114.9  &   5.74  &   66.4 &   5.19 &   96.3 &   5.60 &    91.7 &  5.55 &    86.2 &  5.27 \\ 
Mg\,I &  4702.99 &    4.35  & $-$0.520 &    33.3 &   4.80  &     44.3  &     4.97  &   118.9  &   6.10  &   88.1 &   5.52 &   88.9 &   5.59 &    85.4 &  5.59 &    74.0 &  5.43 \\ 
Mg\,I &  5172.69 &    2.71  & $-$0.380 &   148.9 &   4.76  &    159.3  &     4.62  &\nodata   &\nodata  &\nodata &\nodata &\nodata &\nodata &    85.8 &  5.43 &  \nodata&\nodata\\
...
\enddata
\tablecomments{The table is available in its entirety only in
  electronic format. A portion is shown for guidance regarding its
  form and content.}
\end{deluxetable}


\begin{deluxetable}{lcccccc} 
\tablecolumns{7} 
\tablewidth{0pt} 
\tabletypesize{\small}
\tablecaption{\label{Tab:stellpar} Stellar Parameters}
\tablehead{
\colhead{Star}  &
\colhead{$T_{\rm{eff}}$ } & 
\colhead{$\log (g)$ }    & 
\colhead{$\mbox{[Fe/H]}$ }  & 
\colhead{$v_{\rm{micr}}$ }  \\
\colhead{}&
\colhead{[K]}&
\colhead{[dex]}&\colhead{[dex]}&\colhead{[km\,s$^{-1}$]}}
\startdata
UMa\,II-S1  & 4850 & 1.4 & $-3.10$& 2.0 \\
UMa\,II-NM  & 5200 & 3.5 & $-1.02$& 1.6 \\
UMa\,II-S2  & 4600 & 0.6 & $-3.23$& 2.5 \\
UMa\,II-S3  & 4550 & 1.0 & $-2.34$& 2.2 \\
ComBer-S1      & 4700 & 1.3 & $-2.31$& 2.5 \\
ComBer-S2      & 4600 & 1.4 & $-2.88$& 2.0 \\
ComBer-S3      & 4600 & 1.0 & $-2.53$& 2.2 \\
HD~122563   & 4500 & 0.6 & $-2.77$& 2.5   
\enddata
\tablecomments{Temperatures are rounded to the nearest 10\,K.} 
\end{deluxetable} 

 \begin{deluxetable}{lrrrrrrrrrrrrrrrrrrrrrrrrrrr}
 \tabletypesize{\footnotesize}
 \tabletypesize{\tiny}
 \tablewidth{0pc}
 \tablecaption{\label{abundances_uma} HIRES abundances of the UMa\,II stars}
 \tablehead{
 \colhead{}&
 \multicolumn{5}{c}{UMa\,II-S1}&\colhead{}& \multicolumn{5}{c}{UMa\,II-S2}&\colhead{}& \multicolumn{5}{c}{UMa\,II-S3}\\
 \cline{2-6} \cline{8-12}\cline{14-18}
 \colhead{Species} &
 \colhead{$\lg\epsilon (\mbox{X})$} & \colhead{[X/H]} & \colhead{[X/Fe]}& \colhead{$N$}& \colhead{$\sigma$}&\colhead{}&
 \colhead{$\lg\epsilon (\mbox{X})$} & \colhead{[X/H]} & \colhead{[X/Fe]}&  \colhead{$N$}& \colhead{$\sigma$}&\colhead{}&
 \colhead{$\lg\epsilon (\mbox{X})$} & \colhead{[X/H]} & \colhead{[X/Fe]}&  \colhead{$N$}& \colhead{$\sigma$} }
 \startdata
 C\,    &    6.06 &  $-$2.33 &     0.77 &  2 &     0.20 &&    5.70  & $-$2.69 &    0.54  &  2 & 0.20    &&    5.24 &$-$3.15 & $-$0.81 &   2& 0.20   \\
 O\,I   & $<$7.30 & $<-$1.36 &  $<$1.74 &  1 &  \nodata && $<$7.84  &$<-$0.82 & $<$2.41  &  1 &  \nodata&& $<$6.96 &$<-$1.70& $<$0.64 &   1&\nodata \\
 Na\,I  &    3.06 &  $-$3.11 &  $-$0.01 &  2 &     0.30 &&    2.93  & $-$3.24 & $-$0.01  &  2 & 0.30    &&    3.68 & $-$2.49& $-$0.15 &   2& 0.30   \\
 Mg\,I  &    4.87 &  $-$2.66 &     0.44 &  5 &     0.11 &&    4.77  & $-$2.76 &    0.47  &  5 & 0.15    &&    5.95 & $-$1.58&    0.76 &   4& 0.17   \\
 Al\,I  & $<$5.23 & $<-$1.14 &  $<$1.96 &  1 &  \nodata && $<$5.25  &$<-$1.12 & $<$2.11  &  1 &  \nodata&&    3.70 & $-$2.66& $-$0.34 &   1& 0.30 \\
 Si\,I  & $<$5.77 & $<-$1.74 &  $<$1.36 &  1 &  \nodata && $<$5.74  &$<-$1.77 & $<$1.46  &  1 &  \nodata&&    6.08 & $-$1.43&    0.91 &   1& 0.30 \\
 Ca\,I  &    3.41 &  $-$2.90 &     0.20 &  9 &     0.30 &&    3.46  & $-$2.85 &    0.38  & 12 & 0.19    &&    4.50 & $-$1.81&    0.53 &  22& 0.17   \\
 Sc\,II &    0.20 &  $-$2.85 &     0.25 &  3 &     0.18 && $-$0.30  & $-$3.35 & $-$0.12  &  2 & 0.18    &&    0.67 & $-$2.38& $-$0.04 &   5& 0.15   \\
 Ti\,I  &    2.07 &  $-$2.83 &     0.27 &  6 &     0.24 &&    1.93  & $-$2.97 &    0.26  &  5 & 0.24    &&    2.60 & $-$2.30&    0.04 &  21& 0.10   \\
 Ti\,II &    2.02 &  $-$2.88 &     0.22 & 15 &     0.18 &&    1.80  & $-$3.10 &    0.13  & 16 & 0.23    &&    2.77 & $-$2.13&    0.21 &  27& 0.20   \\
 V\,I   & $<$1.58 & $<-$2.42 &  $<$0.68 &  1 &  \nodata && $<$1.37  &$<-$2.63 & $<$0.60  &  1 &  \nodata&&    1.73 & $-$2.27& $ $0.07 &   2& 0.10   \\
 Cr\,I  &    2.28 &  $-$3.36 &  $-$0.26 &  4 &     0.09 &&    1.91  & $-$3.73 & $-$0.50  &  3 & 0.25    &&    2.98 & $-$2.66& $-$0.32 &  11& 0.13   \\
 Mn\,I  &    2.27 &  $-$3.12  & $-$0.02 &  2 &     0.10 && $<$2.40  &$<-$2.99 & $<$0.24  &  1 &  \nodata&&    2.48 & $-$2.91& $-$0.57 &   3& 0.10   \\
 Fe\,I  &    4.35 &  $-$3.10 &\nodata   & 74 &     0.16 &&    4.22  & $-$3.23 &\nodata   & 64 & 0.16    &&    5.11 & $-$2.34& \nodata & 133& 0.15   \\
 Fe\,II &    4.35 &  $-$3.10 &     0.00 &  7 &     0.13 &&    4.28  & $-$3.17 &    0.06  &  6 & 0.23    &&    5.11 & $-$2.34&    0.00 &  19& 0.16   \\
 Co\,I  & $<$2.77 & $<-$2.15 &  $<$0.95 &  1 &  \nodata && $<$2.59  &$<-$2.33 & $<$0.90  &  1 &  \nodata&&    2.49 & $-$2.43& $-$0.09 &   4& 0.12 \\
 Ni\,I  &    3.44   &$-$2.79 &     0.31 &  3 &     0.40 && $<$3.58  &$<-$2.65 & $<$0.58  &  1 &  \nodata&&    3.87 & $-$2.36& $-$0.02 &  10& 0.16   \\
 Cu\,I  & $<$1.84 & $<-$2.37 &  $<$0.73 &  1 &  \nodata && $<$1.60  &$<-$2.61 & $<$0.62  &  1 &  \nodata&& $<$1.42 &$<-$2.79&$<-$0.45 &   1&\nodata \\
 Zn\,I  &2.35\tablenotemark{a}&$-$2.25&0.85&1&     0.20 &&   $1.42$ & $-$3.18 &  $$0.05  &  1 &  0.20   &&    2.26 & $-$2.34&    0.00 &   1& 0.20   \\
 Sr\,II & $-$0.90 &  $-$3.82 &  $-$0.72 &  1 &     0.30 && $-$1.30  & $-$4.22 & $-$0.99  &  1 &  0.30   && $-$0.90 & $-$3.82& $-$1.48 &  1 & 0.30   \\
 Y\,II  &$<-$1.15 & $<-$3.36 & $<-$0.26 &  1 &  \nodata &&$<-$1.27  &$<-$3.48 &$<-$0.25  &  1 &  \nodata&& $-$1.27 & $-$3.48& $-$1.14 &  1 & 0.15   \\ 
 Zr\,II &$<-0.19$ &$<2.78$  & $<0.32$   & 1  & \nodata  && $<-0.43$  &$<-3.02$&$<0.21$   &  1 & \nodata && $-$0.35 & $-$2.94& $-$0.60 &  1 & 0.30 \\
 Ba\,II & $-$2.27 &  $-$4.44 &  $-$1.34 &  1 &     0.30 && $-$2.20  &$-$4.37  & $-$1.14  &  2 & 0.30    && $-$0.73 & $-$2.90& $-$0.56 &  3 & 0.30   \\
 La\,II &$<-1.54$ &$<-2.67$ & $<0.43$   &$1$ & \nodata  && $<-1.85$ & $<-2.98$&$<0.25$   &$1$ & \nodata && $-$1.49 & $-$2.62& $-$0.28 &  3 & 0.40   \\
 Ce\,II &$<-$1.13 & $<-$2.71 &  $<$0.39 &   1&  \nodata &&$<-$1.49  &$<-$3.07 & $<$0.16  &  1 &  \nodata&&$<-$1.28 &$<-$2.86&$<-$0.52 &  1 &\nodata \\
 Nd\,II &$<-$0.95 & $<-$2.40 &  $<$0.70 &   1&  \nodata &&$<-$1.21  &$<-$2.66 & $<$0.57  &  1 &  \nodata&&$<-$1.43 &$<-$2.88&$<-$0.54 &   1&\nodata \\
 Sm\,II &$<-$0.82 & $<-$1.83 &  $<$1.27 &   1&  \nodata &&$<-$1.53  &$<-$2.54 & $<$0.69  &  1 &  \nodata&&$<-$1.79 &$<-$2.80&$<-$0.46 &   1&\nodata \\
 Eu\,II &$<-$2.00 & $<-$2.52 &  $<$0.58 &   1&  \nodata &&$<-$2.20  &$<-$2.72 & $<$0.51  &  1 &  \nodata&&$<-$2.20 &$<-$2.72&$<-$0.38 &   1&\nodata \\
 \enddata
 \tablecomments{
 [X/Fe] ratios are computed with [Fe\,I/H] abundances
 of the respective stars. Solar abundances have been taken from
 \citet{solar_abund}. See also Table~\ref{Tab:Abundances}. For abundances measured from only one line, we adopt a nominal uncertainty of 0.20\,dex.}
 \tablenotetext{a}{The Zn lines are somewhat distorted; this may lead to an overestimated Zn abundance}
 \end{deluxetable}

 \begin{deluxetable}{lrrrrrrrrrrrrrrrrrrrrrrrrrr}
 \tabletypesize{\footnotesize}
 \tabletypesize{\tiny}
 \tablewidth{0pc}
 \tablecaption{\label{abundances_comber} HIRES abundances of the ComBer stars}
 \tablehead{
 \colhead{}&
 \multicolumn{5}{c}{ComBer-S1}&
 \multicolumn{5}{c}{ComBer-S2}&
 \multicolumn{5}{c}{ComBer-S3}\\
 \cline{2-6} \cline{8-12}\cline{14-18}
 \colhead{Species} &
 \colhead{$\lg\epsilon (\mbox{X})$} & \colhead{[X/H]} & \colhead{[X/Fe]}& \colhead{$N$}& \colhead{$\sigma$}&\colhead{}&
 \colhead{$\lg\epsilon (\mbox{X})$} & \colhead{[X/H]} & \colhead{[X/Fe]}&  \colhead{$N$}& \colhead{$\sigma$}&\colhead{}&
 \colhead{$\lg\epsilon (\mbox{X})$} & \colhead{[X/H]} & \colhead{[X/Fe]}&  \colhead{$N$}& \colhead{$\sigma$}
 }
 \startdata
 C\,    &    6.09 &$-$2.30  &    0.01 & 2 &    0.20 &&    5.60 & $-$2.79 &    0.09 & 2 &   0.20  &&    5.42 & $-$2.97 & $-$0.44 & 2 & 0.20 \\
 O\,I   & $<$7.07 &$<-$1.59 & $<$0.72 & 1 & \nodata && $<$7.34 &$<-$1.32 & $<$1.56 & 1 & \nodata && $<$6.85 &$<-$1.81 & $<$0.72 & 1 & \nodata \\
 Na\,I  &    3.12 & $-$3.05 & $-$0.74 & 2 &    0.30 &&    3.13 & $-$3.04 & $-$0.16 & 1 &    0.30 &&    3.83 & $-$2.34 &    0.19 & 2 & 0.30\\
 Mg\,I  &    5.38 & $-$2.15 &    0.16 & 3 &    0.20 &&    5.63 & $-$1.90 &    0.98 & 3 &    0.10 &&    5.50 & $-$2.03 &    0.50 & 4 & 0.10\\
 Al\,I  & $<$5.25 &$<-$1.12 & $<$1.19 & 1 & \nodata && $<$5.15 &$<-$1.22 & $<$1.66 & 1 & \nodata && $<$5.06 &$<-$1.31 & $<$1.22 & 1 & \nodata \\
 Si\,I  & $<$5.74 &$<-$1.77 & $<$0.54 & 1 & \nodata && $<$5.71 &$<-$1.80 & $<$1.08 & 1 & \nodata && $<$5.48 &$<-$2.03 & $<$0.50 & 1 & \nodata \\
 Ca\,I  &    4.11 & $-$2.20 &    0.11 &18 &    0.18 &&    4.06 & $-$2.25 &    0.63 &17 &    0.21 &&    4.38 & $-$1.93 &    0.60 &22 & 0.15\\
 Sc\,II &    0.32 & $-$2.73 & $-$0.42 & 5 &    0.26 &&    0.72 & $-$2.33 &    0.55 & 4 &    0.19 && $-$0.05 & $-$3.10 & $-$0.57 & 4 & 0.10 \\
 Ti\,I  &    2.41 & $-$2.49 & $-$0.18 &12 &    0.18 &&    2.35 & $-$2.55 &    0.33 &13 &    0.22 &&    2.54 & $-$2.36 &    0.17 &23 & 0.12\\
 Ti\,II &    2.58 & $-$2.32 & $-$0.01 &27 &    0.14 &&    2.42 & $-$2.48 &    0.40 &21 &    0.21 &&    2.66 & $-$2.24 &    0.29 &30 & 0.16\\
 V\,I   & $<$1.18 &$<-$2.82 &$<-$0.51 & 1 & \nodata && $<$1.46 &$<-$2.54 & $<$0.34 & 1 & \nodata && $<$1.10 &$<-$2.90 &$<-$0.37 & 1 & \nodata \\
 Cr\,I  &    3.34 & $-$2.30 &    0.01 &10 &    0.16 &&    2.22 & $-$3.42 & $-$0.54 & 4 &    0.17 &&    2.77 & $-$2.87 & $-$0.34 &11 & 0.21\\
 Mn\,I  & $<$2.40 &$<-$2.99 &$<-$0.68 & 1 & \nodata && $<$2.20 &$<-$3.19 &$<-$0.31 & 1 & \nodata &&    2.20 & $-$3.19 & $-$0.66 & 1 & 0.20 \\ 
 Fe\,I  &    5.14 & $-$2.31 &  \nodata&130&    0.17 &&    4.57 & $-$2.88 & \nodata &98 &    0.14 &&    4.92 & $-$2.53 &  \nodata&146& 0.14\\
 Fe\,II &    5.16 & $-$2.29 &    0.02 &18 &    0.18 &&    4.57 & $-$2.88 &    0.00 & 8 &    0.25 &&    4.87 & $-$2.58 & $-$0.05 &20 & 0.12\\
 Co\,I  & $<$2.69 &$<-$2.23 & $<$0.08 & 1 & \nodata && $<$2.56 &$<-$2.36 & $<$0.52 & 1 & \nodata && $<$2.53 &$<-$2.39 & $<$0.14 & 1 & \nodata \\
 Ni\,I  &    3.36 & $-$2.87 & $-$0.56 & 3 &    0.20 &&    3.43 & $-$2.80 &    0.08 & 5 &    0.22 &&    3.51 & $-$2.72 & $-$0.19 & 9 & 0.12\\
 Cu\,I  & $<$1.78 &$<-$2.43 &$<-$0.12 & 1 & \nodata && $<$1.52 &$<-$2.69 & $<$0.19 & 1 & \nodata && $<$1.44 &$<-$2.77 &$<-$0.24 & 1 & \nodata \\
 Zn\,I  &    2.07 & $-$2.53 & $-$0.22 & 1 &    0.20 &&    1.78 & $-$2.82 &    0.06 & 1 &    0.20 &&    2.17 & $-$2.43 &    0.10 & 1 &  0.20 \\
 Sr\,II & $-$0.80 & $-$3.72 & $-$1.41 & 1 &    0.30 && $-$1.60 & $-$4.52 & $-$1.64 & 1 &    0.30 && $-$1.00 & $-$3.92 & $-$1.39 & 1 &  0.30 \\
 Y\,II  &$<-$1.19 &$<-$3.40 &$<-$1.09 & 1 & \nodata &&$<-$1.17 &$<-$3.38 &$<-$0.50 & 1 & \nodata &&$<-$1.57 &$<-$3.78 &$<-$1.25 & 1 & \nodata \\
 Zr\,II &$<-0.18$ &$<-2.77$ &$<-0.46$ & 1 & \nodata &&  $<0.05$&$<-2.54$ &$<0.34$  & 1 & \nodata &&$<-0.85$ &$<-3.44$ &$<-0.91$ & 1 & \nodata  \\
 Ba\,II & $-$2.47 & $-$4.64 & $-$2.33 & 2 &    0.30 && $-$2.62 & $-$4.79 & $-$1.91 & 2 &    0.30 && $-$1.67 & $-$3.84 & $-$1.31 & 3 & 0.30\\
 La\,II &$<-1.57$ &$<-2.70$ &$<-0.39$ &$1$& \nodata &&$<-1.70$ &$<-2.83$ &$<0.05$  & 1 & \nodata &&$<-2.08$ & $<-3.21$& $<-0.68$& 1 & \nodata\\
 Ce\,II &$<-$1.17 &$<-$2.75 &$<-$0.44 & 1 & \nodata &&$<-$1.34 &$<-$2.92 &$<-$0.04 & 1 & \nodata &&$<-$1.69 &$<-$3.27 &$<-$0.74 & 1 & \nodata \\
 Nd\,II &$<-$0.88 &$<-$2.33 &$<-$0.02 & 1 & \nodata &&$<-$0.99 &$<-$2.44 & $<$0.44 & 1 & \nodata &&$<-$1.32 &$<-$2.77 &$<-$0.24 & 1 & \nodata \\
 Sm\,II &$<-$1.18 &$<-$2.19 & $<$0.12 & 1 & \nodata &&$<-$1.35 &$<-$2.36 & $<$0.52 & 1 & \nodata &&$<-$1.71 &$<-$2.72 &$<-$0.19 & 1 & \nodata \\
 Eu\,II &$<-$1.80 &$<-$2.32 &$<-$0.01 & 1 & \nodata &&$<-$2.40 &$<-$2.92 &$<-$0.04 & 1 & \nodata &&$<-$2.40 &$<-$2.92 &$<-$0.39 & 1 & \nodata \\

 \enddata 

 \tablecomments{ [X/Fe] ratios are computed with [Fe\,I/H] abundances
 of the respective stars. Solar abundances have been taken from
 \citet{solar_abund}. See also Table~\ref{Tab:Abundances}. For
 abundances measured from only one line, we adopt a nominal uncertainty
 of 0.20\,dex.}
 \end{deluxetable}

\begin{deluxetable}{lccrrc} 
\tablecolumns{3} 
\tablewidth{0pc} 
\tablecaption{\label{err} Example Abundance Uncertainties for ComBer-S3}
\tablehead{\colhead{Element}&\colhead{Random}& 
\colhead{$\Delta$\mbox{T$_{\rm eff}$}}&\colhead{$\Delta\log g$}& 
\colhead{$\Delta v_{micr}$}&\colhead{Total}\\ 
\colhead{}&\colhead{uncertainty\tablenotemark{a}}&\colhead{+200\,K}& 
\colhead{$+$0.4\,dex}&\colhead{+0.3\,km\,s$^{-1}$}&\colhead{uncertainty\tablenotemark{b}}} 
\startdata 
 C (CH)&  0.20 & 0.40 &$-$0.10 & $-$0.02 & 0.46 \\ 
 Na\,I &  0.30 & 0.29 &$-$0.17 & $-$0.13 & 0.47 \\ 
 Mg\,I &  0.10 & 0.20 &$-$0.08 & $-$0.06 & 0.24 \\ 
 Ca\,I &  0.15 & 0.14 &$-$0.07 & $-$0.06 & 0.22 \\ 
 Sc\,II&  0.10 & 0.09 &   0.09 & $-$0.06 & 0.15 \\ 
 Ti\,I &  0.12 & 0.28 &$-$0.07 & $-$0.03 & 0.31 \\ 
 Ti\,II&  0.16 & 0.06 &   0.09 & $-$0.08 & 0.21 \\ 
 Cr\,I &  0.21 & 0.28 &$-$0.09 & $-$0.06 & 0.37 \\ 
 Mn\,I &  0.20 & 0.19 &$-$0.06 & $-$0.01 & 0.28 \\
 Fe\,I &  0.14 & 0.25 &$-$0.08 & $-$0.07 & 0.31 \\ 
 Fe\,II&  0.12 & 0.02 &   0.10 & $-$0.06 & 0.17 \\ 
 Ni\,I &  0.12 & 0.22 &$-$0.04 & $-$0.01 & 0.25 \\
 Zn\,I &  0.20 & 0.06 &   0.04 & $-$0.02 & 0.21 \\
 Sr\,II&  0.20 & 0.11 &   0.04 & $-$0.19 & 0.37 \\
 Ba\,II&  0.15 & 0.14 &   0.10 & $-$0.06 & 0.35
\enddata 

\tablenotetext{a}{Standard deviation of individual line abundances (as
given in Tables~\ref{abundances_uma} and \ref{abundances_comber}). For
elements with just one line we adopt a nominal random uncertainty of
0.20\,dex.}

\tablenotetext{b}{Obtained by adding all uncertainties in quadrature.}
\end{deluxetable}

\begin{deluxetable}{llrrrrrrrrr}
\tabletypesize{\footnotesize}
\tabletypesize{\tiny}
\tablewidth{0pc} 
\tablecaption{\label{Tab:Abundances} HIRES abundances of standard star HD~122563}
\tablehead{ 
\colhead{}&\colhead{}&
\multicolumn{5}{c}{HD~122563 -- this study}& \colhead{}& \multicolumn{3}{c}{Aoki et al. (2007)}\\
\cline{3-7}& \cline{8-10}\\
\colhead{Species} & \colhead{$\log\epsilon (\mbox{X})_{\odot}$} &
\colhead{$\log\epsilon (\mbox{X})$} & \colhead{[X/H]} & \colhead{[X/Fe]}& \colhead{$N$}& \colhead{$\sigma$}&\colhead{}&\colhead{$\log\epsilon (\mbox{X})$} & \colhead{[X/H]} & \colhead{[X/Fe]}
}
\startdata          
C\,    & $8.39$ &   5.55 &$-$2.84 &$-$0.07 & 2 & 0.15  &&   5.45 &$-$2.94 &$-$0.35 \\
O\,    & $8.66$ &   6.40 &$-$2.26 &   0.51 & 1 & 0.15  &&   7.03 &$-$1.63 &   1.14 \\
Na\,I  & $6.17$ &   3.41 &$-$2.76 &   0.01 & 2 & 0.10  &&   3.39 &$-$2.78 &$-$0.19 \\
Mg\,I  & $7.53$ &   5.31 &$-$2.22 &   0.55 & 3 & 0.10  &&   5.44 &$-$2.09 &   0.50 \\
Al\,I  &  6.37  &$<$3.87 &$<-$2.50&$<$0.27 & 1 &\nodata&&   3.38 &$-$2.99 &$-$0.40 \\
Si\,I  &  7.51  &   5.42 &$-$2.09 &   0.68 & 1 & 0.15  &&   5.34 &$-$2.17 &   0.41 \\
Ca\,I  & $6.31$ &   3.85 &$-$2.46 &   0.31 &20 & 0.08  &&   3.93 &$-$2.38 &   0.21 \\
Sc\,II & $3.05$ &   0.16 &$-$2.89 &$-$0.12 & 5 & 0.13  &&   0.57 &$-$1.79 &   0.80 \\
Ti\,I  & $4.90$ &   2.27 &$-$2.63 &   0.14 &21 & 0.07  &&   2.42 &$-$2.48 &   0.11 \\
Ti\,II & $4.90$ &   2.40 &$-$2.50 &   0.27 &31 & 0.12  &&   2.58 &$-$2.32 &   0.27 \\
V\,I   &  4.00  &   1.13 &$-$2.87 &$-$0.10 & 1 & 0.15  && \nodata&\nodata&\nodata  \\
Cr\,I  & $5.64$ &   2.56 &$-$3.08 &$-$0.31 &12 & 0.21  &&   2.57 &$-$3.07 &$-$0.48 \\

Mn\,I  & $5.39$ &   2.26 &$-$3.13 &$-$0.36 & 3 & 0.08  &&   2.23 &$-$3.16 &$-$0.57 \\
Fe\,I  & $7.45$ &   4.68 &$-$2.77 &\nodata &141& 0.12  &&   4.86 &$-$2.59 &\nodata \\
Fe\,II & $7.45$ &   4.67 &$-$2.78 &$-$0.01 &20 & 0.10  &&   4.87 &$-$2.58 &   0.01 \\
Co\,I  &  4.92  &   2.32 &$-$2.60 &   0.17 & 1 & 0.15  &&   2.48 &$-$2.44 &   0.15 \\
Ni\,I  & $6.23$ &   3.56 &$-$2.67 &   0.10 &11 & 0.13  &&   3.66 &$-$2.57 &   0.02 \\
Cu\,I  &  4.21  &$<$0.20 &$<-$4.01&$<-$1.24& 1 &\nodata&&\nodata &\nodata&\nodata \\
Zn\,I  & $4.60$ &   1.96 &$-$2.64 &   0.13 & 2 & 0.10  &&   2.08 &$-$2.52 &   0.07 \\
Sr\,II & $2.92$ &$-$0.40 &$-$3.32 &$-$0.55 & 1 & 0.15  &&   0.08 &$-$2.84&$-$0.25 \\
Y\,II  &  2.21  &$-$1.07 &$-$3.28 &$-$0.51 & 2 & 0.07  &&$-$0.93\tablenotemark{a} &$-$3.14&$-$0.37 \\ 
Zr\,II & $2.59$ &$-$0.28 &$-$2.87 &$-$0.11 & 2 & 0.10  &&$-$0.28\tablenotemark{a} &$-$2.87&$-$0.10 \\ 
Ba\,II & $2.17$ &$-$1.77 &$-$3.94 &$-$1.17 & 3 & 0.10  &&$-$1.69 &$-$3.86&$-$1.27 \\
Ce\,II &  1.58  &$-$2.00 &$-$3.58 &$-$0.81 & 1 & 0.15  &&$-$1.83\tablenotemark{a} &$-$3.41&$-$0.64 \\
Nd\,II &  1.45  &$<-$2.22&$<-$3.67&$<-$0.90& 1 &\nodata&&$-$2.01\tablenotemark{a} &$-$3.46&$-$0.69 \\
Sm\,II &  1.01  &$-$2.37 &$-$3.38 &$-$0.61 & 1 & 0.15  &&$-$2.16\tablenotemark{a} &$-$3.17&$-$0.40 \\ 
Eu\,II & $0.52$ &$-$2.46 &$-$2.98 &$-$0.21 & 1 & 0.15  &&$-$2.77\tablenotemark{a} &$-$3.29&$-$0.52 \\
\enddata 
\tablecomments{ [X/Fe] ratios are computed using the [Fe\,I/H]
abundance. Solar abundances have been taken from \citet{solar_abund}.}
\tablenotetext{a}{Neutron-capture abundances are taken from \citet{honda06}}
\end{deluxetable}

\end{document}